\documentclass[fleqn,usenatbib]{mnras}

\usepackage{newtxtext,newtxmath}

\usepackage[T1]{fontenc}

\DeclareRobustCommand{\VAN}[3]{#2}
\let\VANthebibliography\thebibliography
\def\thebibliography{\DeclareRobustCommand{\VAN}[3]{##3}\VANthebibliography}

\usepackage{graphicx}	
\usepackage{amsmath}	

\usepackage{soul}

\newcommand{\fml}{\textsc{FML}}
\newcommand{\mtm}{\textsc{map2map}}
\newcommand{\LCDM}{$\Lambda$CDM}
\newcommand{\pseudo}{\textit{pseudo}$\Lambda$CDM}

\newcommand{\Hrc}{$H_0 r_c$}

\title[A field-level reaction for screening]{A field-level reaction for screened modified gravity}

\author[D. Saadeh et al.]{
Daniela Saadeh,$^{1}$\thanks{E-mail: daniela.saadeh@port.ac.uk (DS)}
Kazuya Koyama,$^{1, 2}$
Xan Morice-Atkinson$^{1}$
\\
$^{1}$Institute of Cosmology \& Gravitation, University of Portsmouth, Dennis Sciama Building,
Burnaby Road, Portsmouth, PO1 3FX, UK \\
$^{2}$ Kavli IPMU (WPI), UTIAS, The University of Tokyo, Kashiwa, Chiba 277-8583, Japan
}

\date{Accepted XXX. Received YYY; in original form ZZZ}

\pubyear{\the\year{}}

\begin{document}
\label{firstpage}
\pagerange{\pageref{firstpage}--\pageref{lastpage}}
\maketitle

\begin{abstract}
We present a field-level reaction framework to emulate the nonlinear effects of screened modified gravity on the cosmic web. This approach is designed to enable field-level inference with data from Stage IV cosmological surveys. Building on the reaction method, which models the nonlinear matter power spectrum in modified gravity as corrections to a ``pseudo'' \LCDM{} cosmology, we extend the method to full field-level predictions by applying it to the output of $N$-body simulations, including both positions and velocities. We focus on modifications to gravity that are scale-independent at the linear level, allowing us to isolate and emulate nonlinear deviations, particularly screening effects. Our neural network predicts the field-level correction (``reaction'') to a \pseudo{} simulation whose linear clustering matches that of the target. The emulator achieves sub-percent accuracy across a broad range of summary statistics, including 0.4\% agreement in the matter power spectrum at scales $k < 1$ Mpc$/h$, and 2\% accuracy in redshift-space distortion multipoles at $k < 0.3$ Mpc$/h$. We also validate the emulator against $N$-body simulations with increased force resolution and time steps, confirming the robustness of its performance. These results demonstrate that our framework is a practical and reliable tool for incorporating screened modified gravity models into field-level cosmological inference, enabling stringent tests of extra fundamental forces at cosmological scales.
\end{abstract}




\section{Introduction}

Four fundamental forces---gravity, electromagnetism, and the strong and weak nuclear forces---underpin all known physical phenomena. However, the possibility of additional, as-yet-undetected forces remains open, and motivates a broad array of experimental and observational efforts. Precise constraints on such hypothetical forces have been established across a wide range of scales, from laboratory experiments to solar system tests and astrophysical observations \citep{Will_Review,Novel_probes,Review_SS_astro_constr,2022PhRvD.106d4040B}. The multi-messenger event GW170817 has confirmed that gravitational waves propagate at the speed of light within parts per $10^{15}$ \citep{GW170817}, placing extremely stringent constraints on the presence of extra forces \citep{Tessa_GW,Vernizzi_GW,Zuma_GW}. However, at cosmological scales, constraints are looser \citep{Novel_probes,Jiamin_review}. Recent analyses of Dark Energy Spectroscopic Instrument (DESI) data \citep{DESI2024} suggest that a simple cosmological constant may not be the best fit to the data, favouring a dynamical model \citep{DESI2024}. These findings prompt renewed interest in alternative dark energy models, particularly those involving additional fundamental forces.

Ongoing experiments such as Euclid\footnote{https://www.euclid-ec.org} and the Vera C. Rubin Observatory\footnote{https://www.lsst.org} allow us to perform much more stringent tests on cosmological scales. By mapping the distribution of dark and luminous matter, these surveys will offer new tests of whether gravity alone governs the growth of cosmic structure, or whether additional forces contribute. Much of the new information from these surveys will reside on nonlinear scales \citep{borg_4}. These scales are particularly important in tests of extra forces, because viable models must be equipped with screening mechanisms to comply with local constraints \citep{koyama,Novel_probes}. In other words, a viable model of a fifth fundamental force must be able to suppress deviations from general relativity where they are known to be small. A handful of screening mechanisms are known: in particular, in chameleon-like screening \citep{chameleon1,chameleon2,symmetron,dilaton}, the strength of the fifth force depends on environmental density, whereas in kinetic screening \citep{Vainshtein,2009PhRvD..79f4036N}, it is suppressed in regimes of high curvature. Theories of extra forces having screening mechanisms include the popular Hu--Sawicki $f(R)$ gravity theory \citep{Hu_Sawicki}, which exhibits chameleon screening, and Dvali--Gabadadze--Porrati (DGP) \citep{DGP} \citep{2009PhRvD..79f4036N}, which employs kinetic screening.

\begin{figure*}
    \centering
    \includegraphics[width=\textwidth]{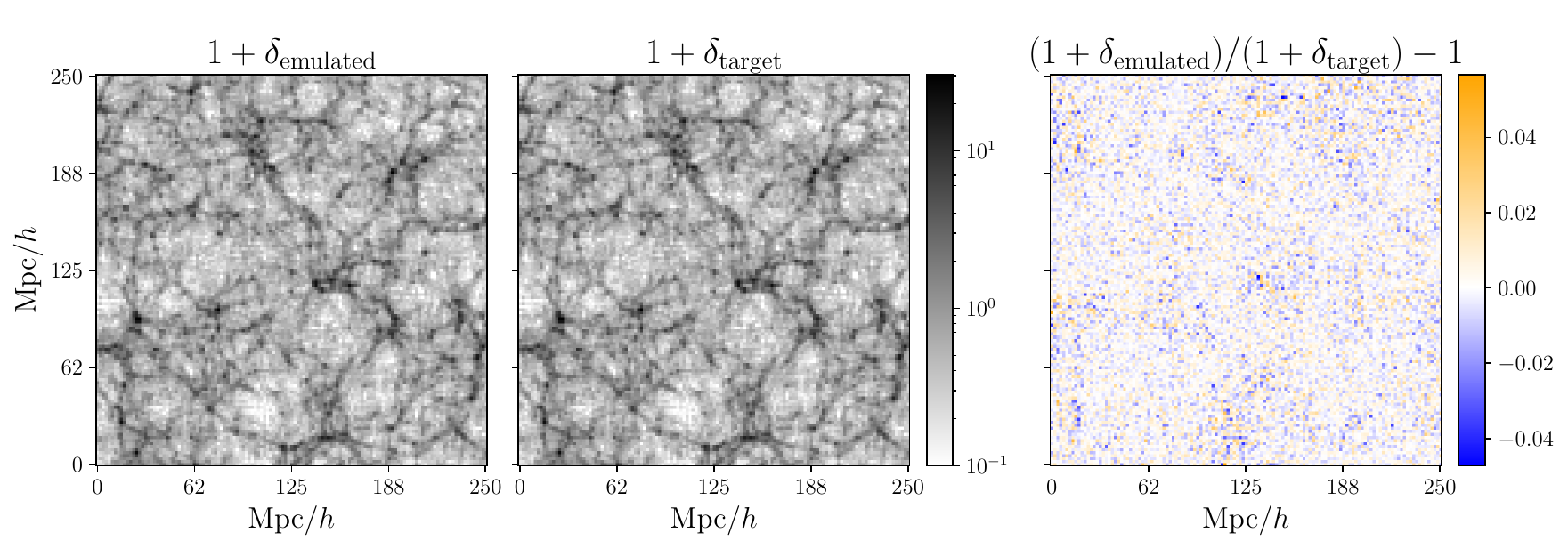}
    \caption{Comparison of the emulator’s output against the target snapshot. The plot shows a slice $250$ Mpc$/h$ $\times$ $250$ Mpc$/h$ wide and $29.3$ Mpc$/h$ deep. In this simulation, run under nDGP gravity, the matter density is $\Omega_M=0.319$, the baryon density $\Omega_b=0.049$, the scalar spectral index $n_s=0.96$, the Hubble parameter today $H_0 = 67$ km/s/Mpc and the amplitude of scalar fluctuations $\sigma_8=0.816$; the strength of modified gravity is given by \Hrc$=1.2$. The left and middle columns show the emulated and target overdensity, whereas the rightmost column shows the relative error on the emulator’s output. This slice is reproduced to $5\%$ at the field level.}
    \label{fig:snapshots}
\end{figure*}

Accurate predictions of models of extra forces with screening would allow us to transform the landscape of constraints at cosmological scales, particularly from the cosmic web. However, detailed modelling at these scales is challenging: perturbative methods break down, and computationally expensive $N$-body simulations become necessary. Incorporating them in the analysis of large cosmological dataset is, therefore, impractical. To preserve accuracy, data at the nonlinear scales must then be removed \citep{DES2023}, forfeiting its strong constraining power on dark energy physics \citep{borg_4}. 

An additional benefit of $N$-body simulations is the potential to move beyond summary statistics. Traditionally, summary statistics like the matter power spectrum are extracted from observations and compared against theoretical predictions. This allows us to average quantities of interest over several sources, making the comparison less sensitive to the specifics of single objects. However, this approach necessarily discards valuable information: analysing all the information at the field level would be more ambitious and stringent. Doing so requires overcoming formidable obstacles: it is necessary to have detailed theoretical predictions at the field level, excellent control of systematics, and survey-specific modelling. Despite these challenges, the payoff is significant: constraints on cosmological parameters and fundamental physics could improve by factors of $\sim$3-5 \citep{nguyen2024information}.

In the past few years, substantial effort has shifted from the powerful -- but lossy -- compression of summary statistics toward full field-level inference -- see, for instance, \citet{BORG,Dvorkin, Lemos, WL, WL2}. One strategy to mitigate the cost of full simulations is emulation -- learning mappings from computationally cheaper quantities to more complex ones. \citet{Angulo_2010} developed a rescaling algorithm working to generate \LCDM\ cosmologies with varied parameters starting from a single simulation. This approach was later extended to modified gravity theories such as $f(R)$ \citep{Mead_2015}. These works (and follow-ups) work by developing prescriptions to transform an $N$-body simulation into another by displacing the particles so that the reconstructed field matches target properties.

Other approaches have sought to model the nonlinear regime agnostically using parameterisations capturing several theories at once. In particular, the \textit{reaction method} \citep{React_1,React_2,React_3,Lombriser_2016} emulates the nonlinear power spectrum in modified gravity by comparing it to a $\Lambda$CDM model with matched linear clustering. The nonlinear spectrum of this reference model is termed the \emph{pseudo-power spectrum}, and the ratio of the target to this pseudo-spectrum is called the \emph{reaction}. This decomposition allows the effort to shift to emulating the reaction, which encapsulates the key nonlinear effects, including screening.

The Artificial Intelligence revolution has offered a separate venue to emulation, with machine learning algorithms showing great potential as emulators. Most of the effort has gone into emulating the power spectrum -- see, for example, \citet{EE2, Bacco, COSMOPOWER, matryoshka} -- which was recently extended to cover modified gravity models \citep{FORGE, eMantis, Fiorini2023}. More recently, machine learning was employed at field level. \citet{map2map} developed a styled neural network transforming snapshots produced under the Zel{'}dovich approximation into the fully nonlinear field evolved in an $N$-body simulations, as predicted by the \textsc{QUIJOTE} simulations \citep{QUIJOTE}. This emulator, called \mtm{}, has also been incorporated into the Bayesian Origin Reconstruction from Galaxies (\textsc{BORG}) algorithm \citep{BORG,borg_1, borg_2, borg_3, borg_4, borg_5}. Building upon that work, we recently developed a field-level emulator for $f(R)$ gravity \citep{our_emulator}, utilising \mtm\ to transform $\Lambda$CDM simulations to $f(R)$ simulations across a range of modified gravity strengths treated as a style parameter. Summary statistics extracted from emulated fields shows excellent agreement with the target simulations: for example, the matter power spectrum was $1\%$ accurate at scales $k < 1$ Mpc$/h$. Independently, \citet{Carolina} used conditional generative adversarial networks to also learn the mapping from \LCDM\ and $f(R)$ gravity, obtaining $5\%$ accuracy up to scales $k < 2$ Mpc$/h$.

In this work, we draw inspiration from the reaction method to develop a \emph{field-level} reaction emulator for screened modified gravity. For simplicity, we will focus on the normal-branch DGP model (nDGP), where linear modifications to gravity reduce to a scale-independent rescaling of the amplitude of scalar fluctuations. This simplifies the setup, allowing us to isolate and emulate the nonlinear effects captured by the reaction. Our emulator takes a $\Lambda$CDM simulation matched to the nDGP target at the linear level, and predicts the corresponding field-level reaction—thus concentrating solely on the nonlinear corrections, including kinetic screening. While we focus on nDGP here, the method is general and applicable to scale-dependent theories as well.

The field-level emulator we present can be paired with \mtm\ to generate full nDGP simulations. By selecting a $\Lambda$CDM amplitude matching the linear behaviour of nDGP, and applying our emulator for the nonlinear corrections, we offer a cost-effective and accurate method for exploring modified gravity effects in the large-scale structure.

This paper is structured as follows. We begin in Section~\ref{sec:method}, where we describe the methodology underlying this work. In particular, we outline the Reaction Method in Section~\ref{fig:reac-meaning}, explaining how we adapt it to the field level. Section~\ref{sec:mapping} details the mapping that our emulator is trained to learn, while Section~\ref{sec:simulations} presents the simulations used in our training, validation, and test sets, including their specifications and parameters. We evaluate the emulator's performance on the test set in Section~\ref{sec:results}. Specifically, Section~\ref{sec:results_summary_statistics} compares the emulated and target snapshots using a broad set of summary statistics, including power spectra, bispectra, redshift-space distortions, and field-level stochasticities. Additionally, in Section~\ref{sec:more-time-steps}, we conduct a robustness test using higher force resolutions than those employed during training. We conclude in Section~\ref{sec:conclusion}.

\section{Methods} \label{sec:method}

In this section, we present the methodology underlying this work. We begin in Section~\ref{sec:reaction} with a brief review of the Reaction Method, which we adapt to the field level. This approach enables us to emulate the effects of modified gravity, particularly screening, by learning corrections to a modified \LCDM{} simulation that shares the same linear clustering as the target. In Section~\ref{sec:mapping}, we describe the specific mapping that the network is trained to learn. Finally, in Section~\ref{sec:simulations}, we provide a detailed account of the simulations used to construct our training, validation, and test datasets, including all relevant cosmological parameters and numerical settings. For further details on the emulator architecture and loss function, we refer the reader to \citet{our_emulator}.

\subsection{The Reaction Method at the field level} \label{sec:reaction}

\begin{figure*}
\centering
\includegraphics[width=\textwidth]{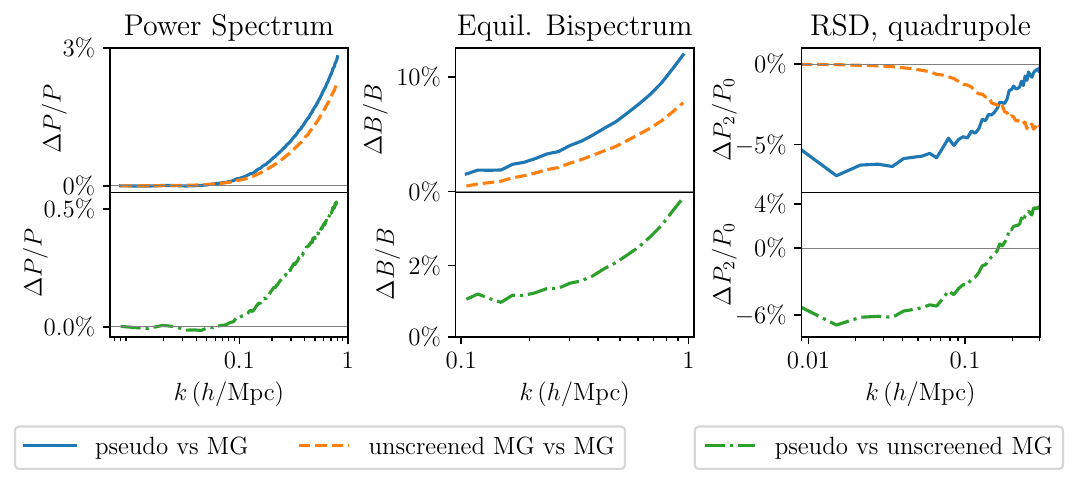}
\caption{Comparison of three simulations used to illustrate the Reaction Method at field level: a target modified gravity simulation (``MG'') with \Hrc$=0.5$, the same MG simulation with screening disabled (``unscreened MG''), and a \pseudo\ simulation (``pseudo'') constructed to match the linear clustering of the target via Eq.~\eqref{eq:pseudo-sigma8} . We consider three observables: the matter power spectrum (left), the equilateral bispectrum (centre), and the quadrupole of the redshift-space distortions (``RSD''; right). In the top panels, we show the fractional deviation of the \pseudo\ (solid blue) and unscreened MG (dashed orange) predictions from the full MG result. The bottom panels show the relative difference between \pseudo\ and unscreened MG simulations (dot-dashed green), highlighting their close agreement in the power spectrum and equilateral bispectrum, and the discrepancy in the RSD case.}
\label{fig:reac-meaning}
\end{figure*}

The Reaction Method, which inspired this work, was introduced by \citet{React_1} and refined in \citet{React_2,React_3}. In this section, we present its essential elements as they apply to our context, referring the interested reader to the cited references for further details.

The Reaction Method aims to emulate the nonlinear matter power spectrum in a target cosmology (typically in modified gravity) starting from that of a \LCDM\ cosmology with the same linear clustering as the target. In nDGP, deviations from \LCDM\ at the linear level are scale-independent: that is, given \LCDM\ and nDGP cosmologies that differ only in the strength of modified gravity \Hrc{}, their linear power spectra differ only by a multiplicative constant \citep{2009PhRvD..79l3512K}. The choice of nDGP therefore allows us to isolate the effects of screening, which is the focus of this work, more easily - however, the method can be easily generalised to other theories.

The (nonlinear) matter power spectrum $P_{\rm pseudo}(k)$---where $k$ is the amplitude of a wave vector---is defined as the matter power spectrum of the \LCDM\ model with the same linear clustering as the target nDGP cosmology. To obtain it, we consider the linear growth equations in nDGP, given by \citep{2009PhRvD..79l3512K}:
\begin{equation}
D^{\prime\prime}(a) + \frac{3}{a} \left( 1 - \frac{\Omega_M}{2 E^2 a^3} \right) D^{\prime}(a) - \frac{3}{2 a^2} \frac{\Omega_M}{E^2 a^3} \left(1 + \frac{1}{3\beta} \right) D(a) = 0,
\end{equation}
where $a$ is the scale factor, $\Omega_M$ is the matter density parameter, $E=H/H_0$, where $H$ is the Hubble parameter in \LCDM{} and the subscript $0$ indicates evaluation at the present time, and:
\begin{equation}
 \beta = 1 + 2 H r_c \left( 1 + \frac{1}{3 H^2}\frac{dH}{dt} \right). 
 \end{equation} 
Note that for \Hrc\ $\rightarrow \infty$, one recovers the linear growth equations of \LCDM. We assume a \LCDM\ background following  \cite{Schmidt:2009sv}.

From the growth factor ratio $D_{\rm nDGP}/D_{\Lambda\mathrm{CDM}}$, one can define the clustering amplitude:
\begin{equation} \label{eq:pseudo-sigma8}
{\sigma_8}^{\rm pseudo} = 
\frac{D_{\rm nDGP}}{D_{\Lambda\mathrm{CDM}}} \sigma_8,
\end{equation}
where $\sigma_8$ is the linear power spectrum amplitude at the scale of 8 Mpc$/h$ in \LCDM, and $h$ is defined by $H_0 \equiv h\times100$ km $\mathrm{s}^{-1},\mathrm{Mpc}^{-1}$.

The pseudo power spectrum $P_{\rm pseudo}(k)$ is then the nonlinear matter power spectrum of a \LCDM\ model with the same cosmological parameters as the target nDGP simulation, except for the clustering amplitude and the strength of modified gravity---set to ${\sigma_8}^{\rm pseudo}$ and \Hrc\ $\rightarrow \infty$, respectively.

The \textit{reaction} is defined as the correction term:
\begin{equation} \label{eq:reaction}
\mathcal{R}(k) \equiv \frac{P_{\rm nDGP}(k)}{P_{\rm pseudo}(k)},
\end{equation}
which encapsulates any residual nonlinear corrections. In the Reaction Method, this term is typically emulated.

In this work, we apply similar ideas to a field-level framework, using Eq.~\eqref{eq:pseudo-sigma8} to define a \pseudo\ simulation, and employing neural networks to emulate the reaction in Eq.\eqref{eq:reaction} at the field level. Neural networks are ideally suited for this task, as they excel at capturing small-scale features within their receptive fields. Available \LCDM\ emulators, such as \textsc{map2map} \citep{map2map}, can then be used to obtain the \pseudo\ simulation using Eq.~\eqref{eq:pseudo-sigma8}.

A central task of the network is to learn the effects of screening, i.e., the ability of viable theories of modified gravity to suppress deviations from general relativity in environments where such deviations must be small. This occurs when the field responsible for the gravitational modification has non-trivial self-interactions. Since screening is intrinsically nonlinear, characterizing it is more challenging than computing linear theory, and for this reason, our emulator is focused on capturing this effect.

We illustrate the concept of starting from a \pseudo\ simulation in Figure~\ref{fig:reac-meaning}. We consider three types of simulations: a target modified gravity simulation (``MG'' in the figure), a second modified gravity simulation with screening disabled (``unscreened MG''), and a \pseudo\ (``pseudo'') simulation. In this example, we adopt a modified gravity strength of \Hrc$=0.5$. We focus on three observables: the matter power spectrum (defined in Eq.~\eqref{eq:power_spectrum}), the bispectrum (defined in Eq.~\eqref{eq:bispectrum}), and the quadrupole moment of the redshift-space distortions (``RSD''; introduced and defined in Sec.~\ref{sec:RSD}). For the bispectrum, we show only the equilateral configuration for conciseness.

The top-left panel shows the deviation of the power spectra of the \pseudo\ (solid blue line) and the unscreened solutions (dashed orange line) from that of the full modified gravity simulation. The bottom-left panel compares the \pseudo\ and the unscreened solutions directly (dot-dashed green line). These panels demonstrate that the \pseudo\ closely resembles the unscreened solution. As screening becomes relevant, however, the \pseudo\ begins to deviate from the full modified gravity simulation, leading to an overestimation of the power spectrum.

This analogy between the \pseudo\ and the unscreened solution also extends to the bispectrum. The middle panel of Figure~\ref{fig:reac-meaning} shows the equilateral bispectrum for the same three models, binned into 20 intervals to mitigate the effects of noise. While the agreement is not as strong as in the case of the power spectrum, the \pseudo\ bispectrum still provides a reasonable approximation to that of the unscreened solution.

The analogy breaks down for other statistics. In particular, the RSD quadrupole, shown in the rightmost panels, exhibits significant discrepancies between the \pseudo\ and even the unscreened solution. The quantity displayed is the quadrupole difference normalized by the monopole, i.e., $\left( {P_{\ell,2}}^{\rm pseudo} - {P_{\ell,2}}^{\rm MG} \right) / {P_{\ell,0}}^{\rm MG}$, and similarly for the bottom panel: this normalization is adopted because the quadrupole crosses zero around $k \approx 0.1$--$0.2$ Mpc$/h$, rendering relative differences difficult to interpret. The \pseudo{} shows deviations of up to 6\% even on the largest scales: this is due to the fact that the growth rate $f = d \ln D/d \ln a$ is different in \pseudo\ and nDGP.

\subsection{Mapping} \label{sec:mapping}

$N$-body simulations model structure formation by evolving initial conditions set in the early Universe all the way up to the present age. This is achieved by initially distributing $N_{\rm part}^3$ particles, each representing a large lump of matter of mass $m$ (of order $\sim 10^{11}-10^{15} M_\odot$), following perturbation theory, and then evolving their positions and velocities under the action of gravity and/or any other forces in the model.

In the Eulerian approach, the large-scale properties of matter are described in terms of fields at space-time locations $(\boldsymbol{x};t)$, which, in $N$-body simulations, are discretized as cells $\{\boldsymbol{x}_{i,j,k}\}$, for $i,j,k\in 0,\cdots,N_{\rm cells}-1 \}$ and time steps $t_l$, for $l\in 0,\cdots,n_{\rm steps}-1$. In particular, key quantities in the Eulerian approach are the density and velocity fields $\rho(\boldsymbol{x},t)$ and $\boldsymbol{v}(\boldsymbol{x},t)$, from which one can define the related density contrast $\delta(\boldsymbol{x},t)$ and momentum fields $\boldsymbol{P}(\boldsymbol{x};t)$ as:
\begin{equation}
\delta(\boldsymbol{x},t) \equiv \frac{\rho(\boldsymbol{x},t)}{\bar{\rho}(t)} -1,
\end{equation}
(where $\bar{\rho}(t)$ is the average density at time $t$), and:
\begin{equation}
\boldsymbol{P}(\boldsymbol{x},t) \equiv m\boldsymbol{v}(\boldsymbol{x};t),
\end{equation}
respectively. Converting the positions and velocities of $N$-body particles into the Eulerian density and velocity fields requires mass assignment schemes that loop over the contributions of all particles following assigned rules: in this work, we use the second-order cloud-in-cell (CIC) scheme, under which each particle is treated as a cube of uniform density and one cell wide, therefore contributing its mass to as many as eight neighbouring voxels.

In the Lagrangian approach, fluid elements are labelled by their starting position $\boldsymbol{q}$ on a uniform grid, and their evolution is characterized in terms of the \textit{displacement} field $\boldsymbol{\Psi}$:
\begin{equation}
\boldsymbol{\Psi}(\boldsymbol{q};t) = \boldsymbol{x}(\boldsymbol{q};t) - \boldsymbol{q},
\end{equation}
representing the change in position of fluid elements. Similarly, one can consider the (Lagrangian) velocity field $\boldsymbol{v}(\boldsymbol{q};t)$.

Following \citet{map2map}, as well as our own recent work \citep{our_emulator}, we train our network to predict the change in the Lagrangian displacement and velocity fields from a \pseudo\ simulation to the target nDGP one:
\begin{equation}
\boldsymbol{\Psi}^{\rm pseudo}(\boldsymbol{q};z=0) \mapsto \boldsymbol{\Psi}^{\rm nDGP}(\boldsymbol{q};z=0),
\end{equation}
and
\begin{equation}
\boldsymbol{v}^{\rm pseudo}(\boldsymbol{q};z=0) \mapsto \boldsymbol{v}^{\rm nDGP}(\boldsymbol{q};z=0).
\end{equation}

A convolutional neural network (CNN) of the kind we employ has a finite receptive field $R$, so by itself it cannot learn correlations on scales $\gg R$. In our setup $R \sim 200$ Mpc$/h$ (from crops of $\sim 1/4$ the boxsize with padding, see \cite{our_emulator}), i.e. sensitivity down to $k \gtrsim 2\pi/R \approx 0.03$ $h/$Mpc.

Crucially, our emulator is designed so that this is not a limitation: we use a field-level reaction (\pseudo{}) step to supply the large-scale modified-gravity response, and the CNN is trained only on the small-scale residuals.
For nDGP specifically, there are no scale-dependent effects on large scales; the scale-independent modifications are captured by the \pseudo{} simulation. For other modified-gravity theories, any large-scale scale dependence is captured by the reaction step as well, so the CNN still only needs to model small-scale corrections within its receptive field.

The only case where the receptive field would matter is if one targeted theories with intrinsically long-range, non-local effects not captured by the reaction mapping; in that case one could increase $R$ (larger crops) or add global-context layers.

\subsection{Simulations} \label{sec:simulations}

Our dataset is made up of 2000 simulations, split 80\%-10\%-10\% as training, validation and test sets, respectively. The \LCDM\ parameters in our simulations are sampled in a Latin hypercube, within the ranges specified in Table~\ref{tab:LH_ranges}; we vary the matter ($\Omega_M$) and baryon ($\Omega_b$) densities, the expansion parameter $h\equiv H_0 / 100 \text{km/s/Mpc}$, where $H_0$ is the Hubble parameter today, the scalar spectral index ($n_s$), and the amplitude of scalar fluctuations on scales of $8$ Mpc$/h$ ($\sigma_8$). The chosen $\Omega_M$ ranges match those of \textsc{Euclid Emulator 2} \citep{EE2}, while the others match those of the \textsc{QUIJOTE} simulations \citep{QUIJOTE}. The ranges in Table~\ref{tab:LH_ranges} are quite broad and span a wide variety of cosmologies.

\begin{table}
    \centering
    \begin{tabular}{c|c|c}
    Parameter      &  min  & max \\
    \hline
    $\Omega_M$     &  0.24 & 0.40 \\
    $\Omega_b$     &  0.03 & 0.07 \\
    $h$            &  0.5  & 0.9  \\
    $n_s$          &  0.8  & 1.2  \\
    $\sigma_8$     &  0.6  & 1    \\
    \hline
    \Hrc        &  0.5  & 7 \\   
    \end{tabular}
   \caption{Cosmological and modified gravity parameters used in our dataset. The $\{ \Omega_M, \Omega_b, h, n_s, \sigma_8 \}$ parameters are sampled using a Latin hypercube within the specified ranges. The modified gravity parameter \Hrc is sampled independently and not as part of the Latin hypercube.}
    \label{tab:LH_ranges}
 \end{table}

For every \LCDM\ simulation, we compute an nDGP simulation with the same ${ \Omega_M, \Omega_b, h, n_s, \sigma_8 }$ parameters and a modified gravity strength \Hrc\ sampled within the range specified in Table~\ref{tab:LH_ranges}---note that \Hrc\ is not sampled in a Latin hypercube with the other parameters. Finally, we compute a \pseudo\ simulation with the same ${ \Omega_M, \Omega_b, h, n_s }$ parameters, but with $\sigma_8$ set according to Eq.~\eqref{eq:pseudo-sigma8}. The Lagrangian displacements and velocities of this simulation form the input of our neural network.

The \Hrc\ range covers very strong (low \Hrc) to very weak (high \Hrc) modifications of gravity. The stronger modifications are especially useful for training the network to learn the effects of modified gravity, while the weaker modifications are more relevant observationally.

We perform all simulations using the \fml{} library\footnote{https://github.com/HAWinther/FML} \citep{FML}, which implements the \textsc{COLA} method \citep{COLA} - see \citet{Winther:2015wla} and \citet{Euclid:2024jej} for a summary and comparison of various simulations for nDGP.  Each simulation has $N=512^3$ particles and a box size of $L=1000$ Mpc$/h$, matching the specifications of the \textsc{QUIJOTE} simulations. All simulations begin at redshift $z=20$, with initial conditions generated using second-order Lagrangian Perturbation Theory (2LPT). The input power spectrum is computed using \textsc{CAMB}\footnote{https://github.com/cmbant/CAMB} \citep{CAMB}. Each simulation evolves through 50 time steps to redshift $z=0$. Following \citet{Albert}, we use a force grid of $N_{\rm force}=3\times N_{\rm part}$ nodes. To reduce the dataset size, simulations are run in single precision, yielding displacement and velocity arrays of approximately 1.6 GB each.

The accuracy of the \fml{} library relative to the emulator should be understood in terms of its strength at reproducing relative differences between models. Previous work has shown that while an individual \fml{} simulation may lack small-scale power compared to high-resolution $N$-body runs, the boost factor — defined as the ratio of the power spectrum in modified gravity relative to a $\Lambda$CDM baseline — is recovered to high accuracy \citep{Gui}. This differential reliability suggests that the library is especially well-suited for emulation strategies like the one adopted here, where the goal is to model screened modified gravity relative to a \LCDM{} reference. In the specific case of nDGP gravity, \cite{Fiorini2023} demonstrated that the boost factor at $z=1$ agrees to within $0.5\%$ between \fml{} and the TreePM code \textsc{Arepo} \citep{Arepo1,Arepo2}, up to $k \sim 5$ $h/$Mpc, when compared on a similar particle resolution. While it remains to be tested whether this level of agreement extends uniformly to higher-order correlators, particularly those sensitive to phases, these results provide evidence that the training data are robust at the differential level. For non-differential comparisons of \textsc{COLA} vs full $N$-body simulations, for different accuracy settings, see \cite{Albert} and \cite{Euclid:2024jej}.

\section{Results} \label{sec:results}

In this section, we evaluate the performance of our emulator on the test set. Figure~\ref{fig:snapshots} illustrates the results, showing side-by-side comparisons of the reconstructed density fields from the emulator and the target simulation (left and middle panels), along with the relative error of the emulator's output (right panel). The figure presents a simulation slice measuring $250$ Mpc$/h$ $\times$ $250$ Mpc$/h$ in area and $29.3$ Mpc$/h$ in depth. The cosmological parameters used are: $\Omega_M = 0.319$, $\Omega_b = 0.049$, $n_s = 0.96$, $H_0 = 67$ km/s/Mpc, $\sigma_8 = 0.816$, and \Hrc$ = 1.2$. In the displayed slice, the emulator recovers the density field with a field-level accuracy of 5\%.

The remainder of this section is structured as follows. First, in Section~\ref{sec:results_summary_statistics}, we present a quantitative evaluation of the emulator's performance using a range of standard summary statistics. These comparisons allow us to assess the accuracy and reliability of the emulator in reproducing the positions and velocities of particles in N-body simulations. To allow comparison of the emulated results against the input \pseudo{}, which already exhibits correct linear clustering, we include the same summary statistics for the input dataset in Appendix~\ref{sec:reac-results} (see in particular Figure~\ref{fig:reac-figure}).

Next, in Section~\ref{sec:more-time-steps}, we perform a complementary analysis by comparing the emulator's output against simulations with higher force resolution and more time steps. Whilst our training, validation, and test datasets are based on the \textsc{COLA} method -- a fast, approximate quasi-N-body method -- we also explore its behavior in the limit of high accuracy. In particular, as the number of time steps  ($n_{\rm steps}$) and the number of grid sites used to compute forces ($N_{\rm force}$) increase, a \textsc{COLA} simulation converges to a full $N$-body simulation having the same number of particles ($N_{\rm part}$), up to some wavenumber. We consider an example simulation where we change  $N_{\rm force}$ from the standard setting $N_{\rm force} = 3 \times N_{\rm part}$, used throughout this work, to a higher-resolution setting of $N_{\rm force} = 5 \times N_{\rm part}$. Simultaneously, we change the number of steps from $n_{\rm steps}=50$ to $n_{\rm steps}=150$. This setup enables a more stringent test of the emulator's robustness and its capacity to generalize to more accurate solvers.

We conclude by presenting the emulator’s speedup relative to an \fml{} run of an nDGP simulation from scratch in Section~\ref{sec:timing}.

\subsection{Performance on summary statistics} \label{sec:results_summary_statistics}

In this Section, we assess the emulator's performance against several summary statistics.

Because our training, validation and test datasets are made up of dark-matter-only simulations, we display any quantity computed from the density and displacement fields only up to scales $k\sim 1\,h/$Mpc to avoid contamination from baryonic effects. Separately, we quote any results calculated from the Eulerian momentum or Lagrangian velocity fields (in particular, the multipoles of the redshift-space distortions) only up to scales $k \sim 0.3\, h/$Mpc: note that the redshift-space distortions smear small-scale nonlinearities to larger scales along the line of sight, due to the “fingers of God” effects \citep{Jackson_1972}. This justifies comparing the redshift-space density power multipoles on much larger scales than the real-space density power spectrum.

Except when computing the power spectra and cross-correlations of vector quantities (i.e. Eqns.~\eqref{eq:vector_power_spectrum}-\eqref{eq:cc_coeff_vec}, for which we use a custom code), we compute all summary statistics showed in this section using the \textsc{Pylians}\footnote{\href{https://pylians3.readthedocs.io}{https://pylians3.readthedocs.io}} library \citep{Pylians}.

The comparison of the same summary statistics extracted from the \pseudo\ simulations is shown in Appendix ~\ref{sec:reac-results}. Although the \pseudo{} simulation served as a useful starting point, our emulator delivers substantial improvements, reducing errors to the subpercent level across several key statistics. In particular, it successfully learns to capture screening effects—a feature entirely absent in the \pseudo{} model. This marks a significant advance over the \pseudo{} approximation, which can exhibit deviations exceeding 10\%.

In all Figures~\ref{fig:density-Pk}-\ref{fig:stochs}, lines are coloured with shades of red proportional to \Hrc\ (as indicated in the colour bar), highlighting the dependence of the summary statistics (and their error) on the strength of the modified gravity parameter/style parameter. Top panels in Figures~\ref{fig:density-Pk}-\ref{fig:RSD} show the ratio between modified gravity and \LCDM{} for the relevant statistic, while bottom panels show the relative emulator error.

\subsubsection{Power spectra}

\begin{figure*}
\centering
\includegraphics[width=\textwidth]{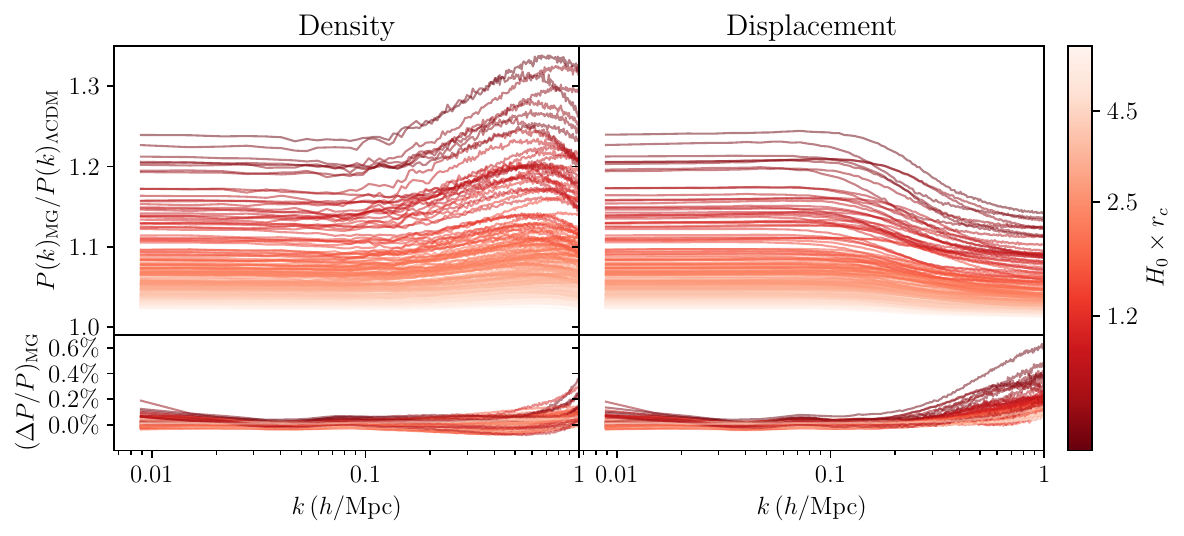}
\caption{\textit{Top panels}: the boost factor \textit{(left)} and the ratio of the power spectra of the displacement field in nDGP and \LCDM\ \textit{(right)}. The power spectrum of the displacement field is defined in Eq.~\ref{eq:vector_power_spectrum}. \textit{Bottom panels}: the fractional error on the density and displacement power spectra as computed from the snapshot produced by our emulator. In all plots, lines are coloured with shades of red proportional to \Hrc{}, as indicated in the colourbar. We find subpercent accuracy for both power spectra at all scales $k<1$ $h/$Mpc.}
\label{fig:density-Pk}
\end{figure*}

\begin{figure*}
\centering
\includegraphics[width=\textwidth]{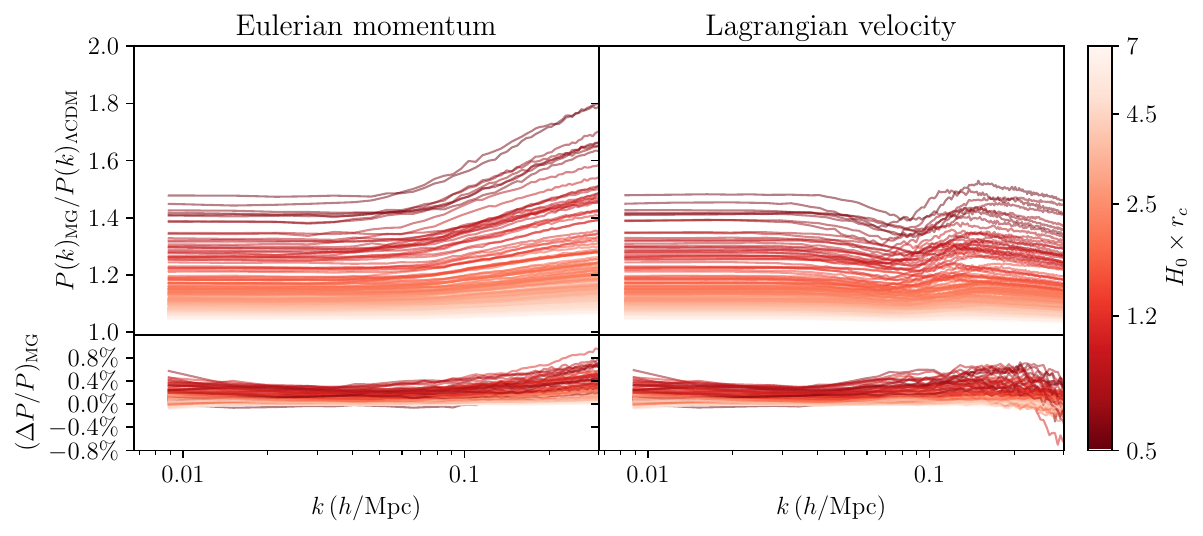}
\caption{\textit{Top panels:} Ratio of the power spectra of the Eulerian momentum \textit{(left)} and Lagrangian velocity fields \textit{(right)}, in nDGP and \LCDM{}. The power spectrum for these fields is defined in Eq.~\eqref{eq:vector_power_spectrum}. \textit{Bottom panels:} the fractional error on the same power spectra, when computed from the snapshot produced by our emulator. We find subpercent agreement at all scales $k < 0.3 \, h/\mathrm{Mpc}$, which are the scales of relevance in the analysis of the redshift-space distortions.}
\label{fig:eul-v-lag-v}
\end{figure*}

Figures~\ref{fig:density-Pk}-\ref{fig:eul-v-lag-v} compare the power spectrum extracted from the emulated and target snapshots, for the density, displacement, Eulerian momentum and Lagrangian velocity fields. The density power spectrum is defined as:
    \begin{equation} \label{eq:power_spectrum}
    	\langle \delta(\boldsymbol{k}_1) \delta(\boldsymbol{k}_2) \rangle = {(2\pi)}^3 \delta_D ( \boldsymbol{k}_1 + \boldsymbol{k}_2 ) P(k),
    \end{equation}
where angular brackets denote ensemble average, $\delta_D$ is the Dirac delta function, and $\boldsymbol{k}_1,\boldsymbol{k}_2$ are Fourier modes.

For vector quantities like the displacement, the Eulerian momentum, and Lagrangian velocity fields, power is summed over spatial dimensions, e.g.:
     \begin{equation} \label{eq:vector_power_spectrum}
    	\langle \boldsymbol{\Psi}(\boldsymbol{k}_1) \cdot \boldsymbol{\Psi} (\boldsymbol{k}_2) \rangle = {(2\pi)}^3 \delta_D( \boldsymbol{k}_1 + \boldsymbol{k}_2) P_{\boldsymbol{\Psi} \boldsymbol{\Psi}}(k),
    \end{equation}
 where the symbol $\cdot$ indicates the Euclidean dot product. As indicated in Sec.~\ref{sec:mapping}, we reconstruct the Eulerian density and momentum from the particles' displacements and velocities using the CIC algorithm.

 The agreement is excellent: all power spectra are recovered to sub-percent accuracy. In particular, the density power spectrum shows a maximum deviation of $0.4\%$, with most predictions well within $0.2\%$ percent error, at scales $k < 1$ $h/$Mpc, and the Lagrangian displacement is recovered to within $0.6\%$ at the same scales. The power spectra of the Eulerian momentum field and the Lagrangian velocity fields are recovered to sub-percent level over scales $k < 0.3$ $h/$Mpc. 

Given this strong agreement, emulator errors at the power spectrum level are likely to be dominated by inaccuracies in the training set, rather than in the machine learning model itself.

\subsubsection{Bispectrum of the density field}
The density bispectrum is defined as:
\begin{equation} \label{eq:bispectrum}
\langle \delta(\boldsymbol{k}_1) \delta(\boldsymbol{k}_2) \delta(\boldsymbol{k}_3) \rangle = {(2\pi)}^3 \delta_D( \boldsymbol{k}_1 + \boldsymbol{k}_2+ \boldsymbol{k}_3) B(k_1, k_2, k_3),
\end{equation}
where the Dirac delta function enforces the condition that the three wavevectors form a closed triangle. This condition arises from statistical homogeneity, and allows the bispectrum to be expressed in terms of two side lengths and the internal angle between them, i.e., $B(k_1,k_2,\theta)$. The bispectrum is the lowest-order correlator sensitive to the phase information of the density field.

It is often useful to work with the \textit{reduced bispectrum}:
\begin{equation}
Q(k_1, k_2, k_3) = \frac{B(k_1, k_2, k_3)}{P(k_1)P(k_2) + P(k_2)P(k_3)+P(k_1)P(k_3)},
\end{equation}
which isolates genuine non-Gaussian contributions not already encoded in the power spectrum.

\begin{figure*}
\centering
\includegraphics[width=\textwidth]{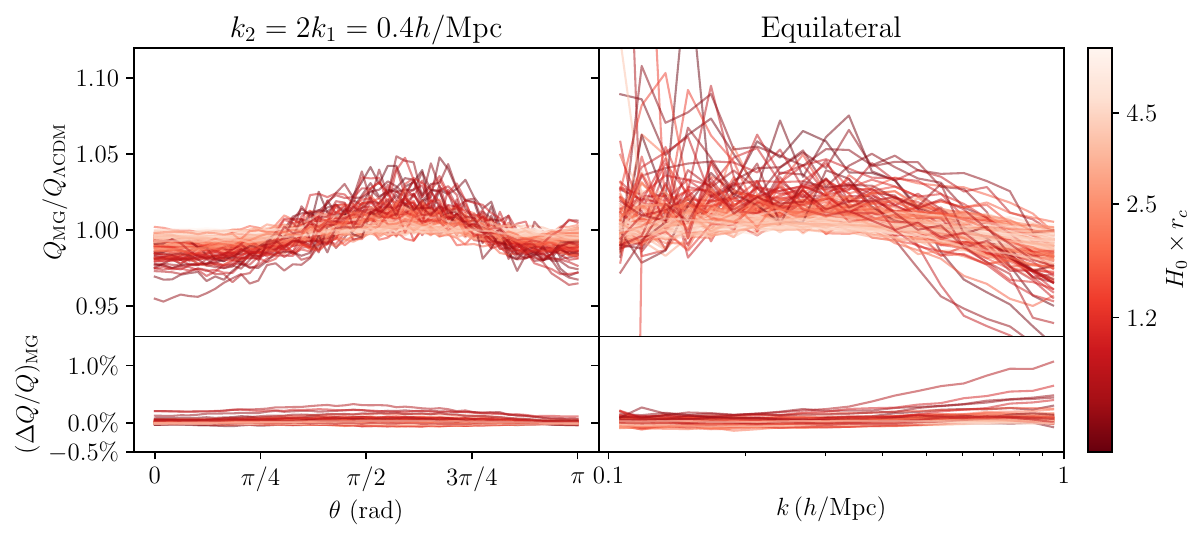}
\caption{\textit{Top panels:} nDGP/\LCDM\ ratio of the reduced bispectra in the elongated configuration $k_2=2k_1=0.4 h$/Mpc \textit{(left)} and in the equilateral configuration \textit{(right)}. \textit{Bottom panels:} the fractional error on the same bispectra, as computed from the output snapshots produced by our emulator. In the equilateral configuration, which includes fewer triangles, we bin both the nDGP/\LCDM{} ratio and the emulator error in 20 intervals, to reduce noise. We have verified that unbinned results follow the same trends. Both configurations show subpercent agreement at all scales $k < 1 h/$Mpc, with nearly all simulations in the test set showing $0.5\%$ accuracy or better.}
\label{fig:bispectrum}
\end{figure*}

In Figure~\ref{fig:bispectrum}, we show the reduced bispectrum in two triangle configurations: the equilateral case, and the elongated configuration with $k_2 = 2 k_1 = 0.4$ $h/\mathrm{Mpc}$. For each case, the top panel displays the ratio $Q_{\rm MG}/Q_{\Lambda\rm CDM}$ as a function of the internal angle $\theta$, while the bottom panel shows the corresponding relative error of the emulator output.

To mitigate noise in the equilateral case, which includes fewer triangles, we bin both the MG/\LCDM{} ratio and the emulator error in 20 intervals. We have verified that unbinned results follow the same trends, confirming that the binned representation in the lower panel of Figure~\ref{fig:bispectrum} is faithful.

The emulator recovers the reduced bispectrum to subpercent accuracy in both configurations, with the majority of simulations achieving errors of $0.5\%$ or better. This confirms that the network captures subtle non-Gaussian signatures introduced by modified gravity, even in highly nonlinear configurations.

\subsubsection{Redshift-space distorsions} \label{sec:RSD}

When observing galaxies falling into gravitational potentials, their peculiar velocities distort the inferred positions along the line of sight in redshift space. These distortions introduce anisotropies in the observed galaxy distribution, known as \textit{redshift-space distortions} (RSD), effectively singling out the line of sight as a preferred direction.

To describe this anisotropy, the redshift-space power spectrum can be expanded in Legendre polynomials $\mathscr{L}$, defined with respect to the line-of-sight unit vector $\hat{\boldsymbol{n}}$:
\begin{equation} \label{eq:RSD_quadrupole}
P(k, \mu) = \sum_{\ell} P_{\ell}(k) \mathscr{L}_{\ell}(\hat{\boldsymbol{k}} \cdot \hat{\boldsymbol{n}}),
\end{equation}
where $\hat{\boldsymbol{k}}$ is the unit wavevector and $\mu = \hat{\boldsymbol{k}} \cdot \hat{\boldsymbol{n}}$.

\begin{figure*}
\centering
\includegraphics[width=\textwidth]{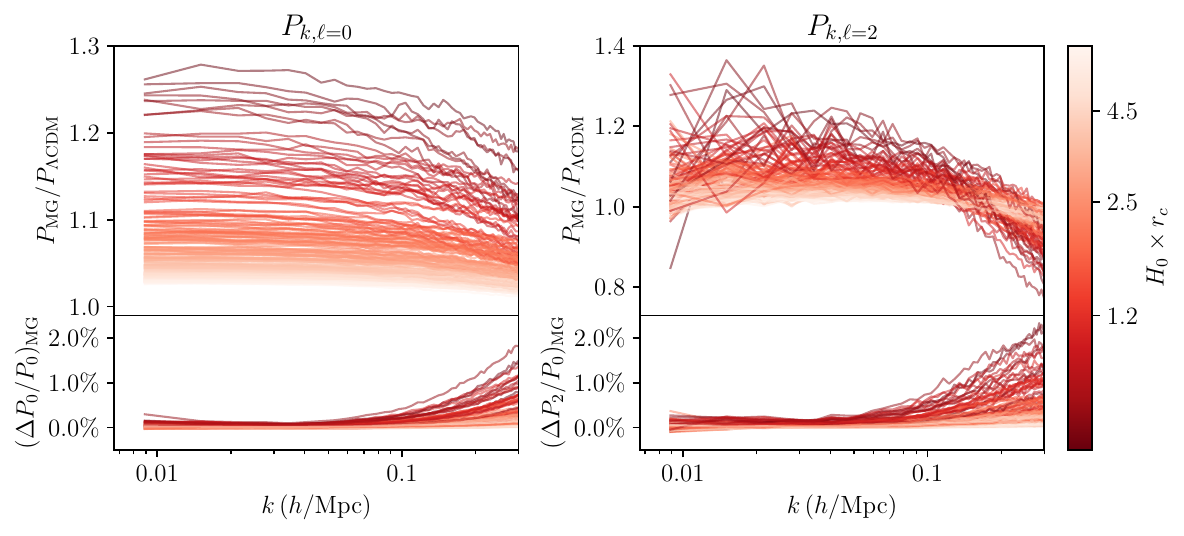}
\caption{\textit{Top panels:} the ratio of the monopole \textit{(left)} and quadrupole \textit{(right)} of the redshift-space distortions between nDGP and \LCDM. \textit{Bottom panels:} fractional error on the monopole \textit{(left)} and quadrupole \text{(right)} as computed from the emulated snapshot (positions and velocities). Because the quadrupole crosses zero at $k \sim 0.1-0.2$ $h$/Mpc, we normalise the error on the quadrupole by the magnitude of the monopole. We find $2\%$ agreement or better in both the monopole and the quadrupole at all scales $k < 0.3 h/$Mpc.}
\label{fig:RSD}
\end{figure*}

In Figure~\ref{fig:RSD}, we evaluate the emulator’s performance on the monopole ($\ell = 0$) and quadrupole ($\ell = 2$) moments of the RSD power spectrum. As in previous figures, the upper panels show the modified gravity signal, expressed as the MG/$\Lambda$CDM ratio, while the lower panels present the emulator error.

Unlike in other statistics, we normalize the absolute error on the quadrupole by the monopole, using $\left( P_{\ell=2,\ \mathrm{out}} - P_{\ell=2,\ \mathrm{true}}\right) /{P_{\ell=0,\ \mathrm{true}}}$, since the quadrupole crosses zero near $k \sim 0.1$--$0.2, h/\mathrm{Mpc}$, making the standard relative error $(P_{\ell=2, \mathrm{out}} - P_{\ell=2, \mathrm{true}}) / P_{\ell=2, \mathrm{true}}$ less informative.

We find that the emulator reproduces the redshift-space multipoles with better than $2\%$ accuracy across all scales $k < 0.3, h/\mathrm{Mpc}$.

\subsubsection{Stochasticities}

The cross-correlation coefficient for the density field is defined as \begin{equation} \label{cc_coeff} r(k) = \frac{ \langle \delta_{\mathrm{out}}(\boldsymbol{k}) \delta_{\mathrm{true}}(\boldsymbol{k}^{\prime})\rangle }{ \sqrt{\langle \delta_{\mathrm{out}}(\boldsymbol{k}) \delta_{\mathrm{out}} (\boldsymbol{k}^{\prime})\rangle} \sqrt{\langle \delta_{\mathrm{true}}(\boldsymbol{k}) \delta_{\mathrm{true}} (\boldsymbol{k}^{\prime}) \rangle} }. \end{equation}

Similarly, for the Eulerian momentum (a vector field), we define:
\begin{equation} \label{eq:cc_coeff_vec}
r(k) = \frac{ \langle \boldsymbol{P}_{\mathrm{out}}(\boldsymbol{k}) \cdot \boldsymbol{P}_{\mathrm{true}}(\boldsymbol{k}^{\prime})\rangle }{ \sqrt{ \langle \boldsymbol{P}_{\mathrm{out}}(\boldsymbol{k}) \cdot \boldsymbol{P}_{\mathrm{out}} (\boldsymbol{k}^{\prime})\rangle} \sqrt{ \langle \boldsymbol{P}_{\mathrm{true}}(\boldsymbol{k}) \cdot \boldsymbol{P}_{\mathrm{true}} (\boldsymbol{k}^{\prime}) \rangle }},
\end{equation}
and analogous definitions apply to the (vector) displacement and Lagrangian velocity fields. In these expressions, the subscripts ``$\mathrm{out}$'' and ``$\mathrm{true}$'' refer to the emulated and reference snapshots, respectively.

The \textit{stochasticity}, given by $1 - r^2$, measures the additional variance in the emulated snapshot that is not captured by the target.

\begin{figure*}
\centering
\includegraphics[width=\textwidth]{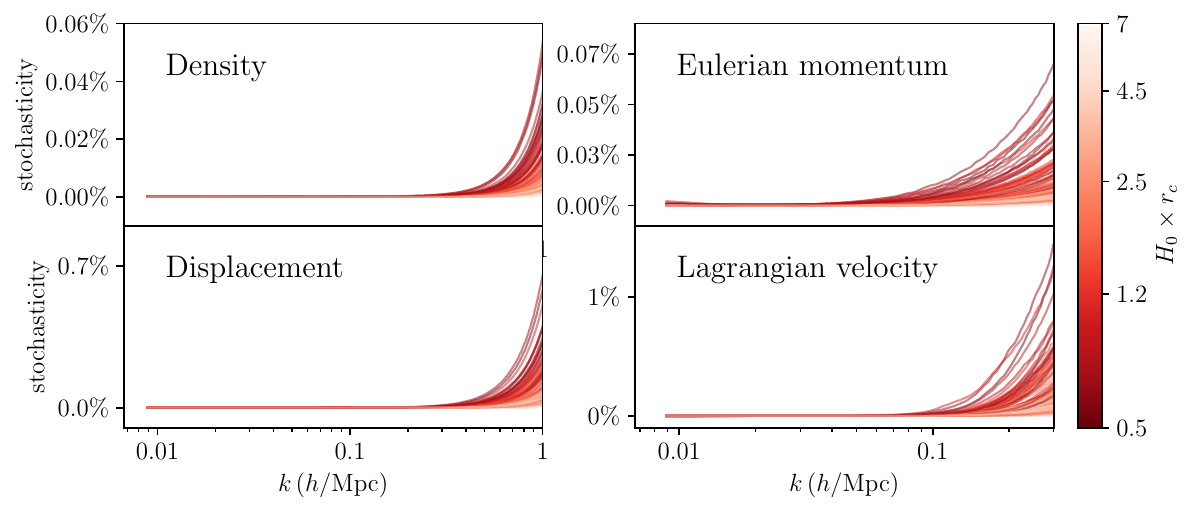}
\caption{Stochasticity of the density \textit{(top left)}, displacement \textit{(bottom left)}, Eulerian momentum \textit{(top right)}, and Lagrangian velocity \textit{(bottom right)} fields. Stochasticity is defined as $1 - r^2$, where $r$ is the cross-correlation coefficient given in Eqs.~\eqref{cc_coeff}--\eqref{eq:cc_coeff_vec}. This quantity is sensitive to phases and measures the variance in the emulated snapshot that is not captured by the target snapshot. We find sub-permille stochasticity in both the density and Eulerian momentum fields at all scales $k < 1\, h/$Mpc and $k < 0.3\, h/$Mpc, respectively. For the displacement field, stochasticity remains below $0.7\%$ for $k < 1\, h/$Mpc. For the Lagrangian velocity field, it stays below $1.5\%$ at $k < 0.3\, h/$Mpc, with most test-set simulations showing stochasticity under $1\%$. These results demonstrate excellent agreement.}
\label{fig:stochs}
\end{figure*}

Figure~\ref{fig:stochs} displays the stochasticities for the density, displacement, Eulerian momentum, and Lagrangian velocity fields. The stochasticity in the emulated density and Eulerian momentum fields remains below the permille level ($\lesssim 0.06$-$0.07\%$) across all scales $k < 1, h/\mathrm{Mpc}$ and $k < 0.3, h/\mathrm{Mpc}$, respectively. The displacement stochasticity remains subpercent at all scales $k < 1, h/\mathrm{Mpc}$, whereas the Lagrangian velocity field has less than $1.5\%$ stochasticity on scales $k < 0.3, h/\mathrm{Mpc}$, with most simulations in the test set staying below 1\%. These results collectively indicate excellent agreement between the emulated and target fields across all relevant scales.

\subsection{Comparison against simulations with higher force resolution} \label{sec:more-time-steps}

In this Section, we perform a complementary analysis by comparing the emulator's output against simulations of higher force resolution than those used in the training set. In general, we do not expect the emulator to perform well on simulations with a different particle resolution—that is, a different value of $L/N_{\rm part}$, where $L$ is the simulation box size and $N_{\rm part}$ is the number of particles. In such cases, we recommend retraining. However, we can still test how well the emulator performs on simulations that maintain the same particle resolution but possess higher \textit{force} resolution and a greater number of time steps.

To clarify this distinction, we remind the reader that our training, validation, and test datasets were generated using the \textsc{COLA} method, as implemented in the \textsc{FML} library. \textsc{COLA} is a fast, approximate quasi-$N$-body method, well-suited to producing a large number of simulations. As the number of time steps and the force resolution increase, \textsc{COLA} solutions converge to a full $N$-body simulation having the same number of particles ($N_{\rm part}$), up to some wavenumber. In particular, \cite{Fiorini2023} showed good agreement at the power spectrum level, up to $k \sim 1$ $h/$Mpc, between \fml{} and the \texttt{TreePM} code \textsc{Arepo} \citep{Arepo1,Arepo2}, when compared against similar particle resolutions.

Comparing against simulations produced with a higher number of time steps and increased force resolution allows us to probe the emulator's performance in the $N$-body limit of \textsc{COLA}. For this, we increase the number of time steps from $n_{\rm steps} = 50$ (our standard setup, see Sec.~\ref{sec:simulations}) to $n_{\rm steps} = 150$; separately, we simultaneously increase the number of grid sites used to compute forces from the standard setting $N_{\rm force} = 3 \times N_{\rm part}$ to a higher-resolution configuration of $N_{\rm force} = 5 \times N_{\rm part}$. We keep the box size unchanged, i.e. $L = 1000$ Mpc$/h$.

\begin{figure*}
    \centering
    \includegraphics[width=\textwidth]{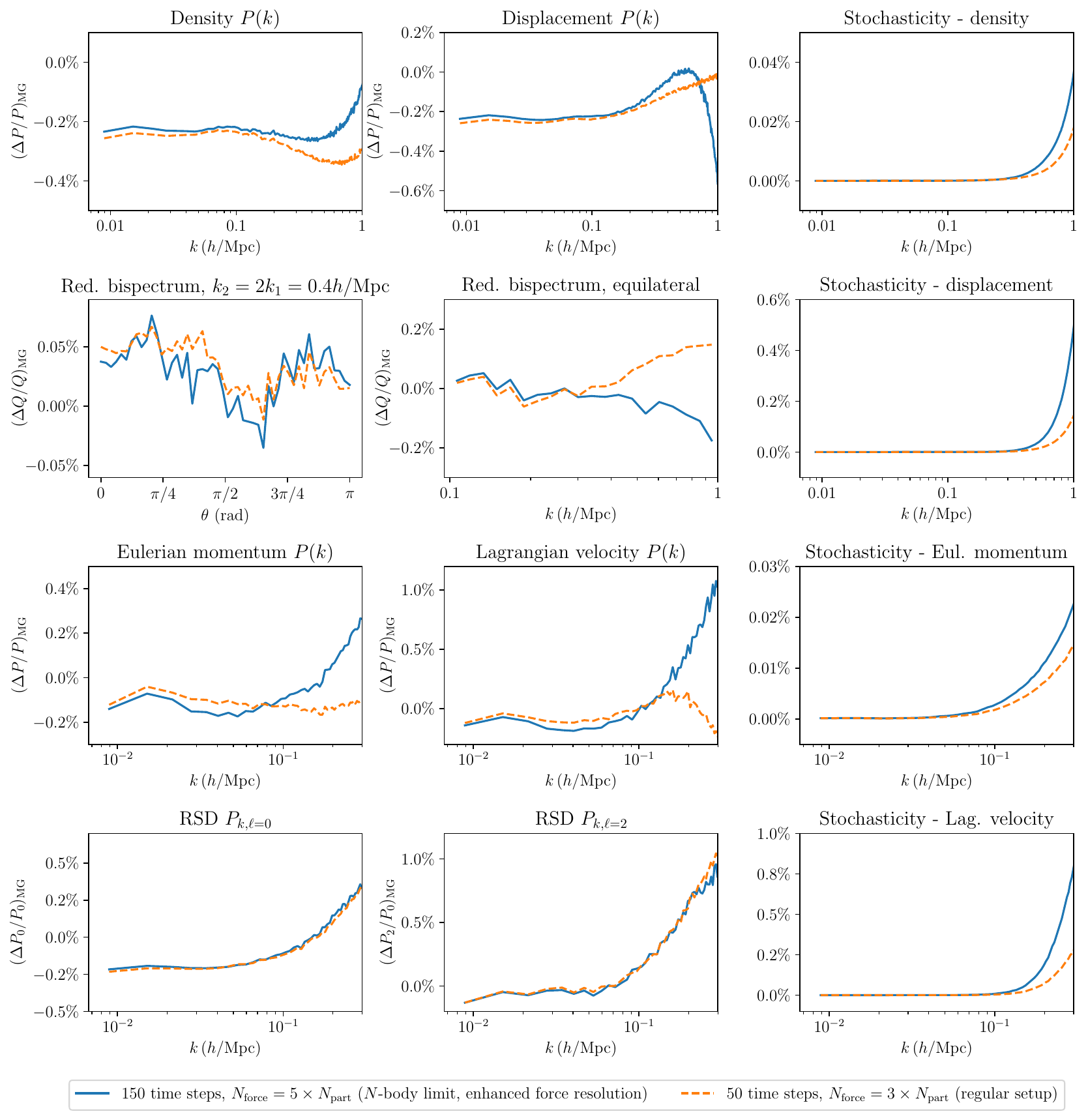}
    \caption{Test of emulator robustness in the $N$-body limit of \textsc{COLA} (the method used to produce the training set), and under increased force resolution. Across the panels, we compare the emulator's performance on two simulations that share identical cosmological and modified gravity parameters (the same as Figure~\ref{fig:snapshots}), but differ in the number of time steps and the force resolution. Specifically, the solid blue line shows results from a higher-resolution simulation approaching the $N$-body limit (150 time steps) with a force resolution of $N_{\rm force} = 5\times N_{\rm part}$, where $N_{\rm part}=512$ is the number of particles. The dashed orange line corresponds to a simulation using our standard setup of 50 time steps and $N_{\rm force} = 3\times N_{\rm part}$. The emulator achieves comparable, sub-percent accuracy in all considered summary statistics across relevant scales.}
    \label{fig:more-time-steps}
\end{figure*}

Figure~\ref{fig:more-time-steps} compares the emulator output on such a simulation, against the corresponding target (blue, solid line). The standard configuration is shown for comparison (orange dashed line). Both simulations share the same cosmological and modified gravity parameters, i.e.: $\Omega_M = 0.319$, $\Omega_b = 0.049$, $n_s = 0.96$, $H_0 = 67$ km/s/Mpc, $\sigma_8 = 0.816$, \Hrc$=1.2$. All other specifications are unchanged compared to the datasets described in Sec.~\ref{sec:simulations}.

The panels display the emulator fractional errors in several key statistics, including the power spectra of the density, displacement, Eulerian momentum, and Lagrangian velocity fields; stochasticities in each of those quantities; the redshift-space monopole and quadrupole; and the reduced bispectrum in both equilateral and elongated triangle configurations.

The emulator maintains sub-percent level accuracy across all of these statistics, in both the standard and higher-resolution simulations. For instance, the density power spectrum error remains below $0.4\%$ across $k < 1, h/\mathrm{Mpc}$, and the Eulerian momentum power spectrum also agrees within $0.4\%$ on scales $k < 0.3, h/\mathrm{Mpc}$. Stochasticities remain at the sub-permille level for the density and Eulerian momentum fields, and below $1\%$ for displacement and velocity fields. 

The reduced bispectrum is also well captured: errors remain sub-permille for the elongated configuration and under permille-level in the equilateral case, demonstrating the emulator's capacity to reproduce higher-order correlations under enhanced force resolution.

These results confirm that our emulator is not only accurate within the force-resolution regime it was trained on, but it can also generalize remarkably well to higher resolutions.

\subsection{Run times} \label{sec:timing}

On a single \textsc{NVIDIA} L40 (40 GB) GPU with 32 CPU cores and 125 GB RAM, the end-to-end per-evaluation time (disk I/O + host/device transfers + GPU forward + writing outputs) is $\approx 26$ s. Following \cite{map2map}, we split each simulation into $64$ crops of $128^3$ particles to fit in GPU memory; the compute-only forward pass is $\approx 0.2\,\mathrm{s}$ per crop ($\approx 12.6\,\mathrm{s}$ total GPU compute per evaluation) when using \texttt{torch.compile} (inductor) and \texttt{bf16} autocast. Inputs/targets reside on a shared Lustre filesystem, and data loading used 4 workers (\texttt{num\_workers=4}, \texttt{pin\_memory=True}). The quoted runtime assumes a precomputed pseudo-$\Lambda$CDM displacement is available (e.g. from the \textsc{map2map} emulator or from a previous \fml{} run) and excludes its generation. Run times from \textsc{map2map} are comparable to ours. A \pseudo{} simulation, run with the \fml{} library for the specifications reported in Sec.~\ref{sec:simulations}, requires $\approx 15$ CPU hours on $16$ CPU-hours on the same node’s host CPUs.

If we include the generation of the \pseudo{} input (with \textsc{map2map}), and assume \(T_{\text{\textsc{map2map}}} \approx T_{\mathrm{pseudo}\to\mathrm{nDGP}} \approx 26\,\mathrm{s}\), we get a total \(T_{\text{\textsc{map2map}}+\mathrm{pseudo}\to\mathrm{nDGP}} \approx 52\,\mathrm{s}\) per realization. By contrast, an end-to-end nDGP simulation run with the \fml{} library (with the specifications listed in Sec.~\ref{sec:simulations}) requires $\approx 40$ CPU-hours on the node’s $32$ host CPUs. This is much faster than a full $N$-body simulation (see \cite{Euclid:2024jej} for a comparison against different $N$-body codes), but still considerably slower than our emulator, which gives a $\approx 2.6\times10^{3}$ CPU-hour$\to$GPU-second speedup; assuming 32 CPU cores, the wall-clock speedup is $\approx 82\times$.


\section{Conclusions} \label{sec:conclusion}

Upcoming surveys like the Vera C. Rubin Observatory and Euclid are set to deliver exquisite maps of the large-scale structure of the Universe. To fully exploit the scientific potential of these datasets, theoretical predictions must reach matching levels of accuracy. For models involving extra fundamental forces, this means accurately capturing the signatures of screening—the mechanism by which the fifth force becomes environment-dependent—a key feature of observationally viable scenarios.

In this work, we introduced a \textit{field-level reaction} framework to emulate the growth of cosmic structure in modified gravity, with particular emphasis on capturing screening effects. The traditional reaction method, applied at the power spectrum level, models modified gravity as a correction to a reference \pseudo{} cosmology—a modified \LCDM{} model with linear clustering similar to that of the target. We extend this idea to the full field level by learning a correction to the positions and velocities of an $N$-body simulation. The reference \pseudo{} cosmology can be efficiently generated using existing \LCDM{} field-level emulators such as \mtm{}.

We validated our emulator against a broad suite of summary statistics, including density and velocity power spectra, the bispectrum, redshift-space multipoles, and field-level stochasticities, at both Eulerian and Lagrangian levels. The model achieves sub-percent accuracy across most observables and relevant scales.

Specifically, the nonlinear matter power spectrum is recovered to better than 0.4\% for $k < 1$ $h/\mathrm{Mpc}$, and the power spectra of velocity and displacement fields remain sub-percent up to $k < 0.3$ $h/\mathrm{Mpc}$. The reduced bispectrum is reproduced at the sub-percent level for both equilateral and elongated triangle configurations. The redshift-space monopole and quadrupole agree with the target within 2\% at the scales relevant for redshift-space distortions, $k < 0.3$ $h/\mathrm{Mpc}$. Field-level stochasticities are well below the permille level for the density and momentum fields, and remain sub-percent and below 1.5\% for the Lagrangian displacement and velocity fields, respectively.

We have also tested the emulator on simulations with increased force resolution, finding consistent performance across all summary statistics. This confirms the robustness of our approach within a range of numerical configurations.

Our field-level reaction framework provides a practical method for incorporating the signatures of extra fundamental forces into the cosmic web, with accurate treatment of screening. As such, it represents a step forward toward field-level inference performed on data from Stage IV experiments, enabling stringent tests of extra fundamental forces at cosmological scales.

\section*{Acknowledgements}

DS, KK and XMA are supported by the STFC grant ST/W001225/1.

Numerical computations were carried out on the \textsc{Sciama} High Performance Computing (HPC) cluster, which is supported by the Institute of Cosmology and Gravitation, the South-East Physics Network (SEPNet) and the University of Portsmouth.

This paper is dedicated to the memory of Graziella Temporin.

For the purpose of open access, we have applied a Creative Commons Attribution (CC BY) licence to any Author Accepted Manuscript version arising.

\section*{Data Availability}
Supporting research data and code are available upon reasonable request from the corresponding author.



\bibliographystyle{mnras}
\bibliography{references} 

\begin{thebibliography}{}
\makeatletter
\relax
\def\mn@urlcharsother{\let\do\@makeother \do\$\do\&\do\#\do\^\do\_\do\%\do\~}
\def\mn@doi{\begingroup\mn@urlcharsother \@ifnextchar [ {\mn@doi@}
  {\mn@doi@[]}}
\def\mn@doi@[#1]#2{\def\@tempa{#1}\ifx\@tempa\@empty \href
  {http://dx.doi.org/#2} {doi:#2}\else \href {http://dx.doi.org/#2} {#1}\fi
  \endgroup}
\def\mn@eprint#1#2{\mn@eprint@#1:#2::\@nil}
\def\mn@eprint@arXiv#1{\href {http://arxiv.org/abs/#1} {{\tt arXiv:#1}}}
\def\mn@eprint@dblp#1{\href {http://dblp.uni-trier.de/rec/bibtex/#1.xml}
  {dblp:#1}}
\def\mn@eprint@#1:#2:#3:#4\@nil{\def\@tempa {#1}\def\@tempb {#2}\def\@tempc
  {#3}\ifx \@tempc \@empty \let \@tempc \@tempb \let \@tempb \@tempa \fi \ifx
  \@tempb \@empty \def\@tempb {arXiv}\fi \@ifundefined
  {mn@eprint@\@tempb}{\@tempb:\@tempc}{\expandafter \expandafter \csname
  mn@eprint@\@tempb\endcsname \expandafter{\@tempc}}}

\bibitem[\protect\citeauthoryear{Adamek et~al.}{Adamek
  et~al.}{2025}]{Euclid:2024jej}
Adamek J.,  et~al., 2025, \mn@doi [Astron. Astrophys.]
  {10.1051/0004-6361/202452180}, 695, A230

\bibitem[\protect\citeauthoryear{{Akhmetzhanova}, {Mishra-Sharma}  \&
  {Dvorkin}}{{Akhmetzhanova} et~al.}{2024}]{Dvorkin}
{Akhmetzhanova} A.,  {Mishra-Sharma} S.,   {Dvorkin} C.,  2024, \mn@doi
  [\mnras] {10.1093/mnras/stad3646}, \href
  {https://ui.adsabs.harvard.edu/abs/2024MNRAS.527.7459A} {527, 7459}

\bibitem[\protect\citeauthoryear{Angulo \& White}{Angulo \&
  White}{2010}]{Angulo_2010}
Angulo R.~E.,  White S. D.~M.,  2010, \mn@doi [Monthly Notices of the Royal
  Astronomical Society] {10.1111/j.1365-2966.2010.16459.x}

\bibitem[\protect\citeauthoryear{{Angulo}, {Zennaro}, {Contreras}, {Aric{\`o}},
  {Pellejero-Iba{\~n}ez}  \& {St{\"u}cker}}{{Angulo} et~al.}{2021}]{Bacco}
{Angulo} R.~E.,  {Zennaro} M.,  {Contreras} S.,  {Aric{\`o}} G.,
  {Pellejero-Iba{\~n}ez} M.,   {St{\"u}cker} J.,  2021, \mn@doi [\mnras]
  {10.1093/mnras/stab2018}, \href
  {https://ui.adsabs.harvard.edu/abs/2021MNRAS.507.5869A} {507, 5869}

\bibitem[\protect\citeauthoryear{{Arnold}, {Leo}  \& {Li}}{{Arnold}
  et~al.}{2019}]{Arepo1}
{Arnold} C.,  {Leo} M.,   {Li} B.,  2019, \mn@doi [Nature Astronomy]
  {10.1038/s41550-019-0823-y}, \href
  {https://ui.adsabs.harvard.edu/abs/2019NatAs...3..945A} {3, 945}

\bibitem[\protect\citeauthoryear{{Arnold}, {Li}, {Giblin}, {Harnois-D{\'e}raps}
   \& {Cai}}{{Arnold} et~al.}{2022}]{FORGE}
{Arnold} C.,  {Li} B.,  {Giblin} B.,  {Harnois-D{\'e}raps} J.,   {Cai} Y.-C.,
  2022, \mn@doi [\mnras] {10.1093/mnras/stac1091}, \href
  {https://ui.adsabs.harvard.edu/abs/2022MNRAS.515.4161A} {515, 4161}

\bibitem[\protect\citeauthoryear{{Bagla}}{{Bagla}}{2002}]{TreePM}
{Bagla} J.~S.,  2002, \mn@doi [Journal of Astrophysics and Astronomy]
  {10.1007/BF02702282}, \href
  {https://ui.adsabs.harvard.edu/abs/2002JApA...23..185B} {23, 185}

\bibitem[\protect\citeauthoryear{{Baker}, {Bellini}, {Ferreira}, {Lagos},
  {Noller}  \& {Sawicki}}{{Baker} et~al.}{2017}]{Tessa_GW}
{Baker} T.,  {Bellini} E.,  {Ferreira} P.~G.,  {Lagos} M.,  {Noller} J.,
  {Sawicki} I.,  2017, \mn@doi [\prl] {10.1103/PhysRevLett.119.251301}, \href
  {https://ui.adsabs.harvard.edu/abs/2017PhRvL.119y1301B} {119, 251301}

\bibitem[\protect\citeauthoryear{{Baker} et~al.,}{{Baker}
  et~al.}{2019}]{Novel_probes}
{Baker} T.,  et~al., 2019, \mn@doi [arXiv e-prints]
  {10.48550/arXiv.1908.03430}, \href
  {https://ui.adsabs.harvard.edu/abs/2019arXiv190803430B} {p. arXiv:1908.03430}

\bibitem[\protect\citeauthoryear{Berger \& Oliger}{Berger \&
  Oliger}{1984}]{AMR}
Berger M.~J.,  Oliger J.,  1984, \mn@doi [Journal of Computational Physics]
  {https://doi.org/10.1016/0021-9991(84)90073-1}, 53, 484

\bibitem[\protect\citeauthoryear{{Boruah} \& {Rozo}}{{Boruah} \&
  {Rozo}}{2024}]{WL2}
{Boruah} S.~S.,  {Rozo} E.,  2024, \mn@doi [\mnras] {10.1093/mnrasl/slad160},
  \href {https://ui.adsabs.harvard.edu/abs/2024MNRAS.527L.162B} {527, L162}

\bibitem[\protect\citeauthoryear{Bose, Cataneo, Tröster, Xia, Heymans  \&
  Lombriser}{Bose et~al.}{2020}]{React_2}
Bose B.,  Cataneo M.,  Tröster T.,  Xia Q.,  Heymans C.,   Lombriser L.,
  2020, \mn@doi [Monthly Notices of the Royal Astronomical Society]
  {10.1093/mnras/staa2696}, 498, 4650–4662

\bibitem[\protect\citeauthoryear{Bose, Tsedrik, Kennedy, Lombriser, Pourtsidou
  \& Taylor}{Bose et~al.}{2023}]{React_3}
Bose B.,  Tsedrik M.,  Kennedy J.,  Lombriser L.,  Pourtsidou A.,   Taylor A.,
  2023, \mn@doi [Monthly Notices of the Royal Astronomical Society]
  {10.1093/mnras/stac3783}, 519, 4780–4800

\bibitem[\protect\citeauthoryear{{Brando}, {Fiorini}, {Koyama}  \&
  {Winther}}{{Brando} et~al.}{2022}]{Gui}
{Brando} G.,  {Fiorini} B.,  {Koyama} K.,   {Winther} H.~A.,  2022, \mn@doi
  [\jcap] {10.1088/1475-7516/2022/09/051}, \href
  {https://ui.adsabs.harvard.edu/abs/2022JCAP...09..051B} {2022, 051}

\bibitem[\protect\citeauthoryear{{Brax}, {van de Bruck}, {Davis}  \&
  {Shaw}}{{Brax} et~al.}{2010}]{dilaton}
{Brax} P.,  {van de Bruck} C.,  {Davis} A.-C.,   {Shaw} D.,  2010, \mn@doi
  [\prd] {10.1103/PhysRevD.82.063519}, \href
  {https://ui.adsabs.harvard.edu/abs/2010PhRvD..82f3519B} {82, 063519}

\bibitem[\protect\citeauthoryear{{Brax}, {Davis}  \& {Elder}}{{Brax}
  et~al.}{2022}]{2022PhRvD.106d4040B}
{Brax} P.,  {Davis} A.-C.,   {Elder} B.,  2022, \mn@doi [\prd]
  {10.1103/PhysRevD.106.044040}, \href
  {https://ui.adsabs.harvard.edu/abs/2022PhRvD.106d4040B} {106, 044040}

\bibitem[\protect\citeauthoryear{Cataneo, Lombriser, Heymans, Mead, Barreira,
  Bose  \& Li}{Cataneo et~al.}{2019}]{React_1}
Cataneo M.,  Lombriser L.,  Heymans C.,  Mead A.~J.,  Barreira A.,  Bose S.,
  Li B.,  2019, \mn@doi [Monthly Notices of the Royal Astronomical Society]
  {10.1093/mnras/stz1836}, 488, 2121–2142

\bibitem[\protect\citeauthoryear{{Creminelli} \& {Vernizzi}}{{Creminelli} \&
  {Vernizzi}}{2017}]{Vernizzi_GW}
{Creminelli} P.,  {Vernizzi} F.,  2017, \mn@doi [arXiv e-prints]
  {10.48550/arXiv.1710.05877}, \href
  {https://ui.adsabs.harvard.edu/abs/2017arXiv171005877C} {p. arXiv:1710.05877}

\bibitem[\protect\citeauthoryear{{DES Collaboration}}{{DES
  Collaboration}}{2023}]{DES2023}
{DES Collaboration} 2023, \mn@doi [\prd] {10.1103/PhysRevD.107.083504}, \href
  {https://ui.adsabs.harvard.edu/abs/2023PhRvD.107h3504A} {107, 083504}

\bibitem[\protect\citeauthoryear{{DESI Collaboration}}{{DESI
  Collaboration}}{2025}]{DESI2024}
{DESI Collaboration} 2025, DESI 2024 VI: Cosmological Constraints from the
  Measurements of Baryon Acoustic Oscillations (\mn@eprint {arXiv}
  {2404.03002}), \mn@doi{10.1088/1475-7516/2025/02/021}, \url
  {https://arxiv.org/abs/2404.03002}

\bibitem[\protect\citeauthoryear{{Doeser}, {Jamieson}, {Stopyra}, {Lavaux},
  {Leclercq}  \& {Jasche}}{{Doeser} et~al.}{2023}]{BORG}
{Doeser} L.,  {Jamieson} D.,  {Stopyra} S.,  {Lavaux} G.,  {Leclercq} F.,
  {Jasche} J.,  2023, \mn@doi [arXiv e-prints] {10.48550/arXiv.2312.09271},
  \href {https://ui.adsabs.harvard.edu/abs/2023arXiv231209271D} {p.
  arXiv:2312.09271}

\bibitem[\protect\citeauthoryear{{Donald-McCann}, {Koyama}  \&
  {Beutler}}{{Donald-McCann} et~al.}{2023}]{matryoshka}
{Donald-McCann} J.,  {Koyama} K.,   {Beutler} F.,  2023, \mn@doi [\mnras]
  {10.1093/mnras/stac3326}, \href
  {https://ui.adsabs.harvard.edu/abs/2023MNRAS.518.3106D} {518, 3106}

\bibitem[\protect\citeauthoryear{{Dvali}, {Gabadadze}  \& {Porrati}}{{Dvali}
  et~al.}{2000}]{DGP}
{Dvali} G.,  {Gabadadze} G.,   {Porrati} M.,  2000, \mn@doi [Physics Letters B]
  {10.1016/S0370-2693(00)00669-9}, \href
  {https://ui.adsabs.harvard.edu/abs/2000PhLB..485..208D} {485, 208}

\bibitem[\protect\citeauthoryear{{Ezquiaga} \& {Zumalac{\'a}rregui}}{{Ezquiaga}
  \& {Zumalac{\'a}rregui}}{2017}]{Zuma_GW}
{Ezquiaga} J.~M.,  {Zumalac{\'a}rregui} M.,  2017, \mn@doi [\prl]
  {10.1103/PhysRevLett.119.251304}, \href
  {https://ui.adsabs.harvard.edu/abs/2017PhRvL.119y1304E} {119, 251304}

\bibitem[\protect\citeauthoryear{Fiorini, Koyama, Izard, Winther, Wright  \&
  Li}{Fiorini et~al.}{2021}]{FML}
Fiorini B.,  Koyama K.,  Izard A.,  Winther H.~A.,  Wright B.~S.,   Li B.,
  2021, \mn@doi [Journal of Cosmology and Astroparticle Physics]
  {10.1088/1475-7516/2021/09/021}, 2021, 021

\bibitem[\protect\citeauthoryear{{Fiorini}, {Koyama}  \& {Baker}}{{Fiorini}
  et~al.}{2023}]{Fiorini2023}
{Fiorini} B.,  {Koyama} K.,   {Baker} T.,  2023, \mn@doi [arXiv e-prints]
  {10.48550/arXiv.2310.05786}, \href
  {https://ui.adsabs.harvard.edu/abs/2023arXiv231005786F} {p. arXiv:2310.05786}

\bibitem[\protect\citeauthoryear{{Gondhalekar}, {Bose}, {Li}  \&
  {Cuesta-Lazaro}}{{Gondhalekar} et~al.}{2025}]{Carolina}
{Gondhalekar} Y.,  {Bose} S.,  {Li} B.,   {Cuesta-Lazaro} C.,  2025, \mn@doi
  [\mnras] {10.1093/mnras/stae2687}, \href
  {https://ui.adsabs.harvard.edu/abs/2025MNRAS.536.1408G} {536, 1408}

\bibitem[\protect\citeauthoryear{{Hern{\'a}ndez-Aguayo}, {Arnold}, {Li}  \&
  {Baugh}}{{Hern{\'a}ndez-Aguayo} et~al.}{2021}]{Arepo2}
{Hern{\'a}ndez-Aguayo} C.,  {Arnold} C.,  {Li} B.,   {Baugh} C.~M.,  2021,
  \mn@doi [\mnras] {10.1093/mnras/stab694}, \href
  {https://ui.adsabs.harvard.edu/abs/2021MNRAS.503.3867H} {503, 3867}

\bibitem[\protect\citeauthoryear{{Hinterbichler} \& {Khoury}}{{Hinterbichler}
  \& {Khoury}}{2010}]{symmetron}
{Hinterbichler} K.,  {Khoury} J.,  2010, \mn@doi [\prl]
  {10.1103/PhysRevLett.104.231301}, \href
  {https://ui.adsabs.harvard.edu/abs/2010PhRvL.104w1301H} {104, 231301}

\bibitem[\protect\citeauthoryear{{Hou}, {Bautista}, {Berti}, {Cuesta-Lazaro},
  {Hern{\'a}ndez-Aguayo}, {Tr{\"o}ster}  \& {Zheng}}{{Hou}
  et~al.}{2023}]{Jiamin_review}
{Hou} J.,  {Bautista} J.,  {Berti} M.,  {Cuesta-Lazaro} C.,
  {Hern{\'a}ndez-Aguayo} C.,  {Tr{\"o}ster} T.,   {Zheng} J.,  2023, \mn@doi
  [Universe] {10.3390/universe9070302}, \href
  {https://ui.adsabs.harvard.edu/abs/2023Univ....9..302H} {9, 302}

\bibitem[\protect\citeauthoryear{{Hu} \& {Sawicki}}{{Hu} \&
  {Sawicki}}{2007}]{Hu_Sawicki}
{Hu} W.,  {Sawicki} I.,  2007, \mn@doi [\prd] {10.1103/PhysRevD.76.064004},
  \href {https://ui.adsabs.harvard.edu/abs/2007PhRvD..76f4004H} {76, 064004}

\bibitem[\protect\citeauthoryear{Izard, Crocce  \& Fosalba}{Izard
  et~al.}{2016}]{Albert}
Izard A.,  Crocce M.,   Fosalba P.,  2016, \mn@doi [Monthly Notices of the
  Royal Astronomical Society] {10.1093/mnras/stw797}, 459, 2327

\bibitem[\protect\citeauthoryear{Jackson}{Jackson}{1972}]{Jackson_1972}
Jackson J.~C.,  1972, \mn@doi [Monthly Notices of the Royal Astronomical
  Society] {10.1093/mnras/156.1.1p}, 156, 1P

\bibitem[\protect\citeauthoryear{{Jamieson}, {Li}, {de Oliveira},
  {Villaescusa-Navarro}, {Ho}  \& {Spergel}}{{Jamieson} et~al.}{2023}]{map2map}
{Jamieson} D.,  {Li} Y.,  {de Oliveira} R.~A.,  {Villaescusa-Navarro} F.,  {Ho}
  S.,   {Spergel} D.~N.,  2023, \mn@doi [\apj] {10.3847/1538-4357/acdb6c},
  \href {https://ui.adsabs.harvard.edu/abs/2023ApJ...952..145J} {952, 145}

\bibitem[\protect\citeauthoryear{{Jasche} \& {Lavaux}}{{Jasche} \&
  {Lavaux}}{2019}]{borg_4}
{Jasche} J.,  {Lavaux} G.,  2019, \mn@doi [\aap] {10.1051/0004-6361/201833710},
  \href {https://ui.adsabs.harvard.edu/abs/2019A&A...625A..64J} {625, A64}

\bibitem[\protect\citeauthoryear{{Jasche} \& {Wandelt}}{{Jasche} \&
  {Wandelt}}{2013}]{borg_1}
{Jasche} J.,  {Wandelt} B.~D.,  2013, \mn@doi [\mnras] {10.1093/mnras/stt449},
  \href {https://ui.adsabs.harvard.edu/abs/2013MNRAS.432..894J} {432, 894}

\bibitem[\protect\citeauthoryear{{Jasche}, {Leclercq}  \& {Wandelt}}{{Jasche}
  et~al.}{2015}]{borg_2}
{Jasche} J.,  {Leclercq} F.,   {Wandelt} B.~D.,  2015, \mn@doi [\jcap]
  {10.1088/1475-7516/2015/01/036}, \href
  {https://ui.adsabs.harvard.edu/abs/2015JCAP...01..036J} {2015, 036}

\bibitem[\protect\citeauthoryear{{Khoury} \& {Weltman}}{{Khoury} \&
  {Weltman}}{2004a}]{chameleon2}
{Khoury} J.,  {Weltman} A.,  2004a, \mn@doi [\prd]
  {10.1103/PhysRevD.69.044026}, \href
  {https://ui.adsabs.harvard.edu/abs/2004PhRvD..69d4026K} {69, 044026}

\bibitem[\protect\citeauthoryear{{Khoury} \& {Weltman}}{{Khoury} \&
  {Weltman}}{2004b}]{chameleon1}
{Khoury} J.,  {Weltman} A.,  2004b, \mn@doi [\prl]
  {10.1103/PhysRevLett.93.171104}, \href
  {https://ui.adsabs.harvard.edu/abs/2004PhRvL..93q1104K} {93, 171104}

\bibitem[\protect\citeauthoryear{Knabenhans et~al.,}{Knabenhans
  et~al.}{2021}]{EE2}
Knabenhans M.,  et~al., 2021, \mn@doi [Monthly Notices of the Royal
  Astronomical Society] {10.1093/mnras/stab1366}, 505, 2840

\bibitem[\protect\citeauthoryear{{Koyama}}{{Koyama}}{2016}]{koyama}
{Koyama} K.,  2016, \mn@doi [Reports on Progress in Physics]
  {10.1088/0034-4885/79/4/046902}, \href
  {https://ui.adsabs.harvard.edu/abs/2016RPPh...79d6902K} {79, 046902}

\bibitem[\protect\citeauthoryear{{Koyama}, {Taruya}  \& {Hiramatsu}}{{Koyama}
  et~al.}{2009}]{2009PhRvD..79l3512K}
{Koyama} K.,  {Taruya} A.,   {Hiramatsu} T.,  2009, \mn@doi [\prd]
  {10.1103/PhysRevD.79.123512}, \href
  {https://ui.adsabs.harvard.edu/abs/2009PhRvD..79l3512K} {79, 123512}

\bibitem[\protect\citeauthoryear{{Lavaux} \& {Jasche}}{{Lavaux} \&
  {Jasche}}{2016}]{borg_3}
{Lavaux} G.,  {Jasche} J.,  2016, \mn@doi [\mnras] {10.1093/mnras/stv2499},
  \href {https://ui.adsabs.harvard.edu/abs/2016MNRAS.455.3169L} {455, 3169}

\bibitem[\protect\citeauthoryear{{Lavaux}, {Jasche}  \& {Leclercq}}{{Lavaux}
  et~al.}{2019}]{borg_5}
{Lavaux} G.,  {Jasche} J.,   {Leclercq} F.,  2019, {Systematic-free inference
  of the cosmic matter density field from SDSS3-BOSS data} (\mn@eprint {arXiv}
  {1909.06396}), \mn@doi{10.48550/arXiv.1909.06396}

\bibitem[\protect\citeauthoryear{{Lemos} et~al.,}{{Lemos} et~al.}{2023}]{Lemos}
{Lemos} P.,  et~al., 2023, in Machine Learning for Astrophysics. p.~18
  (\mn@eprint {arXiv} {2310.15256}), \mn@doi{10.48550/arXiv.2310.15256}

\bibitem[\protect\citeauthoryear{Lewis \& Bridle}{Lewis \& Bridle}{2002}]{CAMB}
Lewis A.,  Bridle S.,  2002, \mn@doi [\prd] {10.1103/PhysRevD.66.103511}, 66,
  103511

\bibitem[\protect\citeauthoryear{{{Ligo Collaboration, Virgo
  Collaboration}}}{{{Ligo Collaboration, Virgo
  Collaboration}}}{2017}]{GW170817}
{{Ligo Collaboration, Virgo Collaboration}} 2017, \mn@doi [\prl]
  {10.1103/PhysRevLett.119.161101}, \href
  {https://ui.adsabs.harvard.edu/abs/2017PhRvL.119p1101A} {119, 161101}

\bibitem[\protect\citeauthoryear{Lombriser}{Lombriser}{2016}]{Lombriser_2016}
Lombriser L.,  2016, \mn@doi [Journal of Cosmology and Astroparticle Physics]
  {10.1088/1475-7516/2016/11/039}, 2016, 039–039

\bibitem[\protect\citeauthoryear{Mead, Peacock, Lombriser  \& Li}{Mead
  et~al.}{2015}]{Mead_2015}
Mead A.~J.,  Peacock J.~A.,  Lombriser L.,   Li B.,  2015, \mn@doi [Monthly
  Notices of the Royal Astronomical Society] {10.1093/mnras/stv1484}, 452,
  4203–4221

\bibitem[\protect\citeauthoryear{Nguyen, Schmidt, Tucci, Reinecke  \&
  Kostić}{Nguyen et~al.}{2024}]{nguyen2024information}
Nguyen N.-M.,  Schmidt F.,  Tucci B.,  Reinecke M.,   Kostić A.,  2024, How
  much information can be extracted from galaxy clustering at the field level?
  (\mn@eprint {arXiv} {2403.03220})

\bibitem[\protect\citeauthoryear{{Nicolis}, {Rattazzi}  \&
  {Trincherini}}{{Nicolis} et~al.}{2009}]{2009PhRvD..79f4036N}
{Nicolis} A.,  {Rattazzi} R.,   {Trincherini} E.,  2009, \mn@doi [\prd]
  {10.1103/PhysRevD.79.064036}, \href
  {https://ui.adsabs.harvard.edu/abs/2009PhRvD..79f4036N} {79, 064036}

\bibitem[\protect\citeauthoryear{{Saadeh}, {Koyama}  \&
  {Morice-Atkinson}}{{Saadeh} et~al.}{2025}]{our_emulator}
{Saadeh} D.,  {Koyama} K.,   {Morice-Atkinson} X.,  2025, \mn@doi [\mnras]
  {10.1093/mnras/stae2807}, \href
  {https://ui.adsabs.harvard.edu/abs/2025MNRAS.537..448S} {537, 448}

\bibitem[\protect\citeauthoryear{{S{\'a}ez-Casares}, {Rasera}  \&
  {Li}}{{S{\'a}ez-Casares} et~al.}{2023}]{eMantis}
{S{\'a}ez-Casares} I.,  {Rasera} Y.,   {Li} B.,  2023, \mn@doi [arXiv e-prints]
  {10.48550/arXiv.2303.08899}, \href
  {https://ui.adsabs.harvard.edu/abs/2023arXiv230308899S} {p. arXiv:2303.08899}

\bibitem[\protect\citeauthoryear{Schmidt}{Schmidt}{2009}]{Schmidt:2009sv}
Schmidt F.,  2009, \mn@doi [Phys. Rev. D] {10.1103/PhysRevD.80.123003}, 80,
  123003

\bibitem[\protect\citeauthoryear{{Spurio Mancini}, {Piras}, {Alsing},
  {Joachimi}  \& {Hobson}}{{Spurio Mancini} et~al.}{2022}]{COSMOPOWER}
{Spurio Mancini} A.,  {Piras} D.,  {Alsing} J.,  {Joachimi} B.,   {Hobson}
  M.~P.,  2022, \mn@doi [\mnras] {10.1093/mnras/stac064}, \href
  {https://ui.adsabs.harvard.edu/abs/2022MNRAS.511.1771S} {511, 1771}

\bibitem[\protect\citeauthoryear{Tassev, Zaldarriaga  \& Eisenstein}{Tassev
  et~al.}{2013}]{COLA}
Tassev S.,  Zaldarriaga M.,   Eisenstein D.~J.,  2013, \mn@doi [Journal of
  Cosmology and Astroparticle Physics] {10.1088/1475-7516/2013/06/036}, 2013,
  036

\bibitem[\protect\citeauthoryear{{Vainshtein}}{{Vainshtein}}{1972}]{Vainshtein}
{Vainshtein} A.~I.,  1972, \mn@doi [Physics Letters B]
  {10.1016/0370-2693(72)90147-5}, \href
  {https://ui.adsabs.harvard.edu/abs/1972PhLB...39..393V} {39, 393}

\bibitem[\protect\citeauthoryear{{Vardanyan} \& {Bartlett}}{{Vardanyan} \&
  {Bartlett}}{2023}]{Review_SS_astro_constr}
{Vardanyan} V.,  {Bartlett} D.~J.,  2023, \mn@doi [Universe]
  {10.3390/universe9070340}, \href
  {https://ui.adsabs.harvard.edu/abs/2023Univ....9..340V} {9, 340}

\bibitem[\protect\citeauthoryear{{Villaescusa-Navarro}}{{Villaescusa-Navarro}}{2018}]{Pylians}
{Villaescusa-Navarro} F.,  2018, {Pylians: Python libraries for the analysis of
  numerical simulations}, Astrophysics Source Code Library, record
  ascl:1811.008 (\mn@eprint {ascl} {1811.008})

\bibitem[\protect\citeauthoryear{{Villaescusa-Navarro}
  et~al.,}{{Villaescusa-Navarro} et~al.}{2020}]{QUIJOTE}
{Villaescusa-Navarro} F.,  et~al., 2020, \mn@doi [\apjs]
  {10.3847/1538-4365/ab9d82}, \href
  {https://ui.adsabs.harvard.edu/abs/2020ApJS..250....2V} {250, 2}

\bibitem[\protect\citeauthoryear{{Will}}{{Will}}{2014}]{Will_Review}
{Will} C.~M.,  2014, \mn@doi [Living Reviews in Relativity]
  {10.12942/lrr-2014-4}, \href
  {https://ui.adsabs.harvard.edu/abs/2014LRR....17....4W} {17, 4}

\bibitem[\protect\citeauthoryear{Winther et~al.}{Winther
  et~al.}{2015}]{Winther:2015wla}
Winther H.~A.,  et~al., 2015, \mn@doi [Mon. Not. Roy. Astron. Soc.]
  {10.1093/mnras/stv2253}, 454, 4208

\bibitem[\protect\citeauthoryear{{Zhou}, {Li}, {Dodelson}  \&
  {Mandelbaum}}{{Zhou} et~al.}{2023}]{WL}
{Zhou} A.~J.,  {Li} X.,  {Dodelson} S.,   {Mandelbaum} R.,  2023, \mn@doi
  [arXiv e-prints] {10.48550/arXiv.2312.08934}, \href
  {https://ui.adsabs.harvard.edu/abs/2023arXiv231208934Z} {p. arXiv:2312.08934}

\makeatother
\end{thebibliography}




\appendix

\section{Comparison between emulated and \pseudo{} results} \label{sec:reac-results}

\begin{figure*}
\centering
\includegraphics[width=\textwidth]{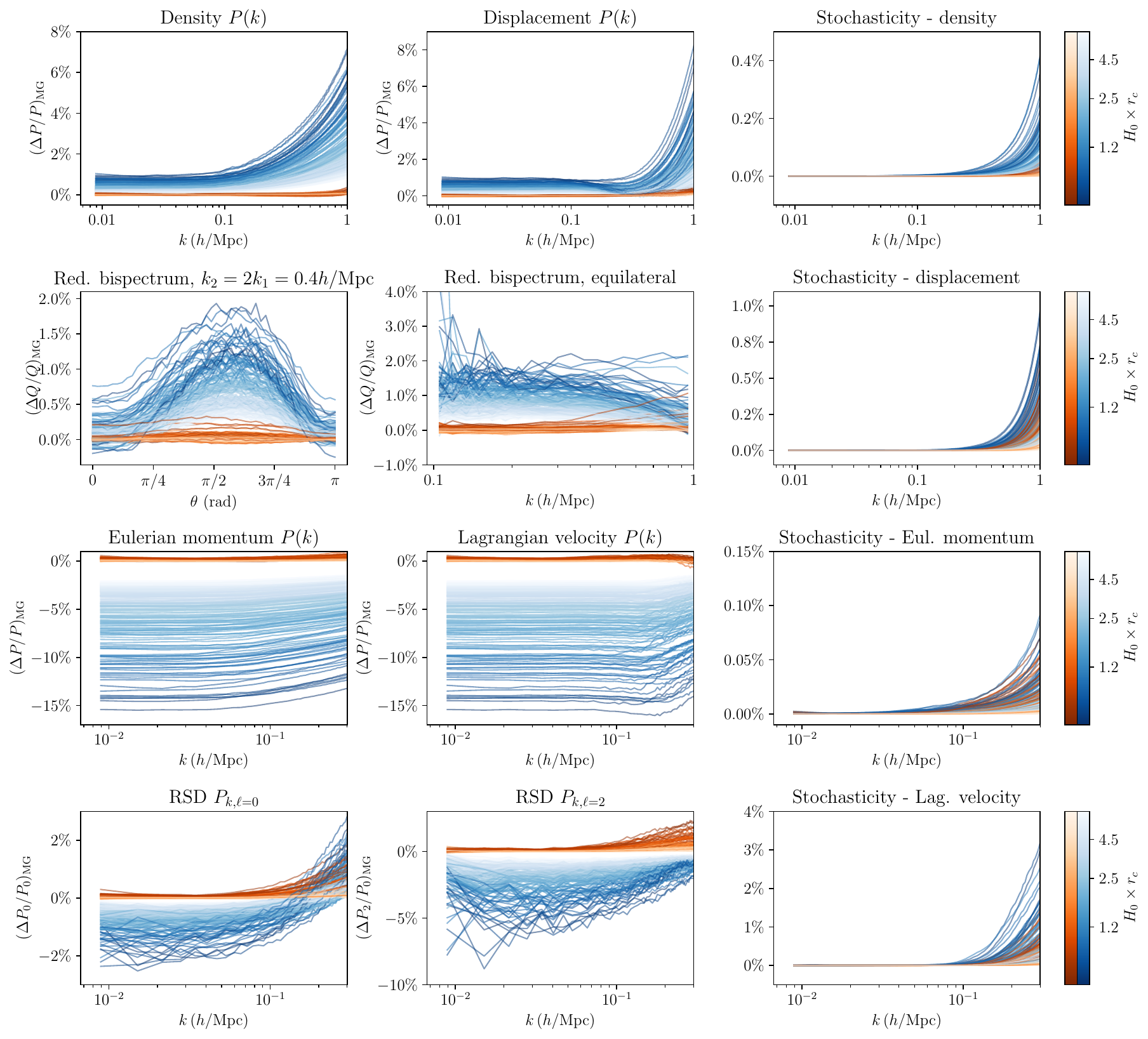}
\caption{Comparison between emulator outputs and \pseudo{} simulations, evaluated using the summary statistics defined in Sec.\ref{sec:results_summary_statistics}. Shades of orange (left colour bar) denote the strength of modified gravity in the emulated snapshots, while shades of blue (right colour bar) indicate the modified gravity strength that the \pseudo{} simulation is constructed to approximate at the level of linear clustering. Large-scale discrepancies in the density and displacement fields for the \pseudo{} are mainly due to imperfect matching of $\sigma_8$ via Eq.\eqref{eq:pseudo-sigma8}. Across all observables, the emulator shows markedly improved agreement with the target simulations, highlighting its capacity to reproduce screening effects that are absent in the \pseudo{}.}
\label{fig:reac-figure}
\end{figure*}

In this Appendix, we compare the emulator's performance to that of the \pseudo{} simulations. As discussed in Sec.\ref{sec:reaction}, the \pseudo{} shares several similarities with an unscreened solution, though this analogy does not carry across all summary statistics. Figure~\ref{fig:reac-figure} presents this comparison, for all the statistics discussed in Sec.~\ref{sec:results_summary_statistics}.

In the Figure, shades of orange (left colour bar) indicate the strength of modified gravity in the emulated snapshots, whilst shades of blue (right colour bar) represent the modified gravity strength that the \pseudo{} simulation is constructed to approximate at the level of linear clustering. Discrepancies at the large scales in the \pseudo{} density and displacement  are primarily due to residual differences in matching $\sigma_8$ via Eq.~\eqref{eq:pseudo-sigma8}. Overall, the emulator significantly outperforms the \pseudo{}, especially in its ability to capture the nonlinear effects of screening.

\bsp	
\label{lastpage}
\end{document}